\documentclass[preprint,12pt,3p]{elsarticle}

\usepackage{amssymb}
\usepackage{float}
\usepackage{caption}
\usepackage{subcaption}
\usepackage{graphicx}
\usepackage{fixltx2e}
\usepackage{amsmath}
\usepackage{mathtools}
\usepackage[makeroom]{cancel}
\usepackage{epstopdf}
\usepackage{gensymb}
\usepackage[]{algorithm2e}
\usepackage[justification=centering]{caption}

\graphicspath{{Images/}}

\journal{Reliability Engineering and System Safety}

\begin{document}

\begin{frontmatter}

\title{Global Sensitivity Analysis via Multi-Fidelity Polynomial Chaos Expansion}

\author[label1,label2]{Pramudita Satria Palar}
\ead{pramsp@ftmd.itb.ac.id, palar@edge.ifs.tohoku.ac.jp}
\address[label1]{Bandung Institute of Technology, Jl. Ganesha No. 10, Bandung, Indonesia}
\address[label2]{Tohoku University, Sendai, Miyagi Prefecture 980-8577, Japan}

\author[label1]{Lavi Rizki Zuhal}

\author[label2]{Koji Shimoyama}

\author[label3]{Takeshi Tsuchiya}
\address[label3]{University of Tokyo, Tokyo, 113-8656, Japan}

\begin{abstract}
The presence of uncertainties are inevitable in engineering design and analysis, where failure in understanding their effects might lead to the structural or functional failure of the systems. The role of global sensitivity analysis in this aspect is to quantify and rank the effects of input random variables and their combinations to the variance of the random output. In problems where the use of expensive computer simulations are required, metamodels are widely used to speed up the process of global sensitivity analysis. In this paper, a multi-fidelity framework for global sensitivity analysis using polynomial chaos expansion (PCE) is presented. The goal is to accelerate the computation of Sobol sensitivity indices when the deterministic simulation is expensive and simulations with multiple levels of fidelity are available. This is especially useful in cases where a partial differential equation solver computer code is utilized to solve engineering problems. The multi-fidelity PCE is constructed by combining the low-fidelity and correction PCE. Following this step, the Sobol indices are computed using this combined PCE. The PCE coefficients for both low-fidelity and correction PCE are computed with spectral projection technique and sparse grid integration. In order to demonstrate the capability of the proposed method for sensitivity analysis, several simulations are conducted.  On the aerodynamic example, the multi-fidelity approach is able to obtain an accurate value of Sobol indices with 36.66\% computational cost compared to the standard single-fidelity PCE for a nearly similar accuracy. 

\end{abstract}


\end{frontmatter}


\section{Introduction}
Computational methods are widely deployed to predict the behavior and to compute the output of interests (e.g., stress distribution or force) of physical or engineering systems. In engineering design and analysis, computational partial differential equation (PDE) solvers are routinely employed to aid and enhance such processes. However, due to the existence of uncertainties in the system, deterministic analysis might result in a failure to understand the true probabilistic nature of the system. An example of this is in the field of aerospace engineering, where manufacturing error and environmental uncertainties such as gust, turbulence, and change in design conditions perturb the nominal condition of the aircraft. Understanding the effect of the uncertainties to the system is then a vital task in many engineering and scientific problems. Depending on the nature of the uncertainties, uncertainties can be categorized as either aleatory or epistemic uncertainties. While aleatoric uncertainties are irreducible and inherent to the system, epistemic uncertainties are caused by our lack of knowledge on the system being investigated. Aleatoric uncertainties can be conveniently expressed by probability theory using specific measures such as statistical moments.

Sensitivity analysis (SA) in probabilistic modeling plays an important role in understanding the impact of input random variables and their combinations to the model output. By performing SA, the contribution of the input random variables to the model output can be ranked, thus giving the analysts information of which variables are the most and least responsible. This is very useful in UQ, where the dimensionality of the random variable can be high and it is desired to reduce the complexity beforehand. Moreover, sensitivity information can provide important knowledge about the physics of the system being investigated. In engineering analysis, numerical evaluation of PDE is often required in order to obtain necessary information for SA. In the field of fluid and solid mechanics, the model responses are typically evaluated using computational fluid dynamics (CFD) or finite element methods (FEM), respectively. 

Based on the range of the domain to be analyzed, SA is commonly classified into two groups~\cite{saltelli2000sensitivity}:
\begin{itemize}
 \item \textit{local sensitivity analysis}, which studies the local impact of the input parameters on the model output and typically provides the partial derivatives of the output to the input parameters.
 \item \textit{global sensitivity analysis}, which studies the uncertainties in the output due to changes of the input parameters over the entire domain.
\end{itemize}


Based on how it exploits the model response, SA can be further categorized into~\cite{saltelli2000sensitivity}:

\begin{itemize}
 \item \textit{regression based methods}, which perform linear regression to the relationship between the model input and output. These methods are limited to cases with a linear or almost linear relationship but are inadequate in cases with highly non-linear relationship.
 \item \textit{variance-based methods}, which decompose the variance of the output into the sum of contributions of each input variables and their combination. In statistic literature, variance based methods are widely known as ANOVA (ANalysis Of VAriance)~\cite{efron1981jackknife}
\end{itemize}

In this paper, we have interest in global SA using variance-based methods since our main focus is on UQ, although the methodology itself can be used for other applications such as optimization. More specifically, this paper focuses on Sobol sensitivity analysis method~\cite{sobol1990sensitivity}.

Monte Carlo simulation (MCS) is the most conventional method to perform SA~\cite{sobol1990sensitivity}. The advantage of MCS for SA is that it is straightforward and very simple to be implemented.  However, it is infeasible to obtain accurate results using MCS if a demanding computational simulation is used. To cope with this, more advanced techniques have to be employed. SA using cheap analytical replacement of the true function, or a metamodel, is one way to handle this. Metamodeling techniques such as radial basis function~\cite{park1991universal}, Kriging~\cite{krige1951statistical, matheron1963principles, cressie2015statistics, sacks1989design, gandin1965objective}, probabilistic collocation~\cite{loeven2007probabilistic}, and polynomial chaos expansion (PCE)~\cite{knio2001stochastic, ghiocel2002stochastic, xiu2002wiener, ghanem2003stochastic} can be employed for SA purpose.

In the closely related field of structural reliability analysis, metamodeling techniques have been widely employed to accelerate the computation of failure probability~\cite{das2000cumulative, papadrakakis2002reliability, kaymaz2005application, dubourg2011metamodel, bichon2011efficient,  balesdent2013kriging}. Compared to other metamodels, Kriging metamodel is particularly useful for reliability analysis due to the direct availability of error estimation that allows one to deploy active learning strategy~\cite{bichon2008efficient, echard2011ak, huang2016assessing} (i.e., Kriging with adaptive sampling). Active learning works by adding more samples so as to increase the accuracy of the metamodel near the region of interest, that is, the limit state. Besides Kriging, PCE has also been employed for structural reliability analysis purpose~\cite{choi2004structural, berveiller2006stochastic, hu2011adaptive, sudret2013response}. Recently, a combination of PCE and Kriging within the framework of universal Kriging and active learning was developed to handle rare event estimation problem~\cite{schobi2016rare}.

Metamodels in UQ and SA are required to be globally accurate for a precise estimation of sensitivity indices. Although not in the context of global SA, the use of Kriging models with adaptive sampling to progressively refine the global accuracy has been explored by some researchers within the context of UQ~\cite{bilionis2012multi, shimoyama2013dynamic}. Dwight and Han proposed an adaptive sampling technique for UQ by using a criterion based on the product of the Kriging uncertainty and the probability density function of the random inputs~\cite{dwight2009efficient}, which is useful for cases with non-uniform distributions. Another alternative is to construct local surrogate models such as Dutch intrapolation~\cite{rumpfkeil2011dynamic} and multivariate interpolation and regression~\cite{wang2010high, boopathy2014unified} in order to provide the error measure. A recent survey of adaptive sampling for Kriging and other surrogate models for various applications can be found in Liu et al.~\cite{liu2017survey}. Other than Kriging, polynomial-based metamodels, especially PCE, are highly useful and competitive to the Kriging model when the quantities of interest are the statistical moments and sensitivity indices.

The advantage of PCE for UQ and SA purpose is that the calculation of Sobol sensitivity indices can be directly performed as the post-processing step~\cite{sudret2008global, crestaux2009polynomial}.  This is in contrast to other metamodelling techniques where MCS still needs to be performed on the analytical metamodel to obtain the Sobol indices. The approximation quality of the metamodel can be further enhanced by including gradient information. Some methods that have been reformulated in order to include gradient information are gradient-enhanced Kriging~\cite{morris1993bayesian, liu2003development, dwight2009efficient, de2014improvements, de2015exploiting}, compressive sensing-based PCE~\cite{peng2016polynomial}, and sparse grid method~\cite{de2015gradient}. However, it is worth noting that obtaining gradient information is a tedious task and is not always possible for many applications. In this paper, we assume that gradient information is not available.

The PCE method is based on homogeneous chaos expansion, which itself is based on the seminal work of Wiener~\cite{wiener1938homogeneous}. The earliest version of PCE computes the coefficients by using Galerkin minimization. This intrusive scheme needs modification of the governing equation and the simulation code in order to perform the UQ procedure~\cite{ghanem2003stochastic}. The difficulty with the intrusive scheme is that the derivation of the modified governing equation might be very complex and highly time-consuming. In this paper, the method of interest is non-intrusive PCE~\cite{ghanem2003stochastic, ghiocel2002stochastic}, that allows the use of legacy code since it can be treated as black box simulation. Based on how it estimates the PCE coefficients, non-intrusive PCE can be classified into two categories:

\begin{enumerate}
	\item \textit{Spectral projection}, which estimates the coefficients by exploiting the orthogonality of the polynomial and quadrature~\cite{knio2001stochastic, ghiocel2002stochastic}.
	\item \textit{Regression}, which estimates the coefficients by using least-square minimization~\cite{choi2004polynomial, berveiller2006stochastic}.
\end{enumerate}

Note that the definition of regression-based PCE is different from the regression-based SA. The definition of the former is that a regression-based technique is used to build the PCE metamodel, while the definition of the latter is the utilization of linear approximation within SA framework. 

PCE has been developed and implemented as a tool for SA since the work of Sudret~\cite{sudret2008global}. Further simulations are not necessary because the Sobol indices in PCE are obtained in the post-processing phase~\cite{sudret2008global, crestaux2009polynomial} by exploiting the orthogonality of the polynomial bases. PCE is now a widely used approach in the field of UQ due to its strong mathematical basis~\cite{eldred2009comparison}. For regression-based PCE, sparse-PCE based on least-angle-regression is an efficient method to obtain the Sobol indices without the need to determine the polynomial bases a priori~\cite{blatman2010efficient, blatman2011adaptive}. Another alternative for sparse PCE is using a compressive sensing-based technique~\cite{doostan2011non, jakeman2015enhancing} that employs orthogonal matching pursuit to scan the most influential polynomial bases. While for the spectral-projection based PCE, the sparse grid interpolation technique~\cite{smolyak1960interpolation, gerstner1998numerical, bungartz2004sparse, garcke2012sparse} can be employed to reduce the number of collocation points in high-dimensionality for SA purpose~\cite{smolyak1963quadrature, xiu2005high, constantine2012sparse, buzzard2012global, buzzard2011variance}. Recent advances in this field include the use of PCE for multivariate SA~\cite{garcia2014global}. Our interest in this paper is to obtain Sobol sensitivity indices for global SA purpose by using polynomial chaos expansion (PCE).

In some applications, simulations with multiple levels of fidelity are available. The fidelity here is defined as the measure of how accurate the model approximates the reality. The fidelity level is typically categorized into high- and low-fidelity, where the categorization of the model into high- or low-fidelity depends on the case being investigated. The high-fidelity (HF) simulation is the most accurate but is typically more computationally demanding than the low-fidelity (LF) simulation. On the other hand, the LF simulation is less accurate but many evaluations could be performed since it is cheaper to evaluate. For example, a PDE-solver with a very fine and coarse mesh can be treated as the HF and LF simulation, respectively. An approach that utilizes simulations with multiple levels of accuracy is commonly called multi-fidelity (MF) method. The common technique to apply MF simulations is to use LF simulations to capture the response trend and employ HF simulations to correct the response. 

Multi-fidelity simulation is commonly used to aid and accelerate the optimization process. Co-Kriging~\cite{kennedy2000predicting, forrester2007multi} and space-mapping~\cite{bandler2004space, shah2015multi} are two examples of surrogate-based methods that rely on MF simulations; mainly for optimization purposes. Multi-fidelity techniques for UQ purpose recently appeared in literature. Among the first is the multi-level Monte Carlo (MLMC) method which was firstly developed in the context of solving stochastic PDEs problem~\cite{giles2008multilevel, barth2011multi} and was further developed to handle general black-box problems~\cite{ng2014multifidelity}. The Kriging method is also applicable for UQ purpose~\cite{de2015uncertainty}. Recently, a non-intrusive MF-PCE based technique to solve UQ problems was developed. The method works by combining the LF and correction PCE into a single MF-PCE~\cite{ng2012multifidelity}. The coefficients of the LF and correction PCE can be obtained by employing spectral projection~\cite{ng2012multifidelity} or regression-based technique~\cite{palar2015decomposition, palar2016multi}. Another similar technique is the MF stochastic collocation that relies on Lagrange-polynomial interpolation~\cite{narayan2014stochastic}. Although the use of MF simulations is common to be found in optimization and, recently, UQ literatures, its application to SA is still rare and scarce. One example of works in MF SA is the application of Co-Kriging for SA of a spherical tank under internal pressure~\cite{le2014bayesian}.

To the best of our knowledge, there is still no research that investigates the capability of MF-PCE for SA purpose. Such work is important to further enhance the capability of MF-PCE for goals other than standard UQ. Furthermore, it has a potential to reduce the computational cost to perform SA, which is desirable in applications that involve expensive simulations. In this paper, we investigate and demonstrate the usefulness and capabilities of MF-PCE to perform SA when simulations with different level of fidelity are available.

The rest of this paper is structured as follows: The explanation of the general framework of SA and the methodology of MF-PCE are given in Section~\ref{sec:2}. The numerical studies on artificial and real-world test problems we undertook are discussed in Section~\ref{sec:3}. Finally, conclusions are drawn and future work is discussed in Section~\ref{sec:4}.

\section{Methodology}
\label{sec:2}
\subsection{Global SA}
\subsubsection{Probabilistic formulation of the problem}
We first consider a physical model with $n$ input parameters, with the relationship between the input vector $\boldsymbol{x}$ and the scalar output $y$ is defined by 
\begin{equation}
	y = f(\boldsymbol{x}), 
\end{equation}
where $\boldsymbol{x}=\{x_{1},\ldots,x_{n}\}^{\text{T}}\in\mathbb{R}^{n}$ and $n\geq1$ is the vector of the input variables. Because our interest in this paper is for UQ, the input vector is assumed as random variables $\boldsymbol{\xi}=\{\xi_{1},\ldots,\xi_{n} \}^{\text{T}}$ with joint probability density function (PDF) $\rho(\boldsymbol{\xi})$. Due to the assumption of independent components of $\boldsymbol{\xi}$, one gets $\rho(\boldsymbol{\xi}) = \prod_{i=1}^{n}\rho_{\xi_{i}}(\xi_{i})$ where $\rho_{\xi_{i}}(\xi_{i})$ is the marginal PDF of $\xi_{i}$. The random output $Y$ is then defined by

\begin{equation}
  Y = f(\boldsymbol{\xi}).
\end{equation}

\subsubsection{Sobol decomposition}
The model response $Y$ can be decomposed into summands of its main effects and interactions (called Sobol decomposition~\cite{sobol1990sensitivity}), reads as
\begin{equation}
 Y = f(\boldsymbol{\xi})=f_{0}+\sum_{i=1}^{n}f_{i}(\xi_{i})+\sum_{1\leq i<j \leq n} f_{i,j}(\xi_{i},\xi_{j})+\ldots+f_{1,2,\ldots,n}(\xi_{1},\ldots,\xi_{n}),
\end{equation}
where $f_{0}$ is a constant and is the mean value of the function defined by
\begin{equation}
 f_{0}=\int_{\boldsymbol{\Omega}}f(\boldsymbol{\xi})\rho(\boldsymbol{\xi})\mathrm{d}\boldsymbol{\xi},
\end{equation}
where $\boldsymbol{\Omega}$ is the support of the PDF. A property of the Sobol decomposition is that its summand satisfies 
\begin{equation}
\label{eq:sobol3}
\begin{aligned}
 \int_{\boldsymbol{\Omega}_{\xi_{k}}}f_{i_{1},\ldots,i_{s}}(\xi_{i_{1}},\ldots,\xi_{i_{s}}) \rho_{\xi_{k}}(\xi_{k}) \mathrm{d}\xi_{k}=0\\
 \text{for } 1\leq i_{1} <\ldots<i_{s}\leq n, k\in\{i_{1},\ldots,i_{s}\},
\end{aligned}
\end{equation}
where $\boldsymbol{\Omega}_{\xi_{k}}$ is the support of the PDF. As a consequence of Eq.~\ref{eq:sobol3}, the summands except $f_{0}$ are mutually orthogonal in the following sense:

\begin{equation}
\label{eq:sobol4}
\begin{aligned}
\int_{\boldsymbol{\Omega}}f_{i_{1},\ldots,i_{s}}(\xi_{i_{1}},\ldots,\xi_{i_{s}}) f_{j_{1},\ldots,j_{t}}(\xi_{j_{1}},\ldots,\xi_{j_{t}})\mathrm{d}\boldsymbol{\xi}=0,\\
\text{for } \{i_{1},\ldots,i_{s} \} \neq \{j_{1},\ldots,j_{t} \}.
\end{aligned}
\end{equation}
Based on this derivation, we can then derive the formula of sensitivity indices.

\subsubsection{Sensitivity indices}
As a consequence of the orthogonality property shown in Eq.~\ref{eq:sobol4}, the total variance $D$ of the model response $f(\boldsymbol{\xi})$ can then be decomposed as
\begin{equation}
 \mathbb{V}[f(\boldsymbol{\xi})]=D = \sum_{i=1}^{n}D_{i}+\sum_{1\leq i < j \leq n}D_{i,j}+\ldots+D_{1,2,\ldots,n},
\end{equation}
where the $D_{i_{1},\ldots,i_{s}}$ are the \textit{partial variances} and defined as
\begin{equation}
 D_{i_{1},\ldots,i_{s}}=\mathbb{V}[f_{i_{1},\ldots,i_{s}} (\xi_{i_{1}},\ldots,\xi_{i_{s}})].
\end{equation}
We can then define the Sobol indices 
\begin{equation}
 S_{i_{1},\ldots,i_{s}}=D_{i_{1},\ldots,i_{s}}/D,
\end{equation}
where they satisfy the following relationship:
\begin{equation}
\sum_{i=1}^{n}S_{i}+\sum_{1\leq i<j\leq n} S_{i,j}+\ldots+S_{1,2,\ldots,n}=1.
\end{equation}
This means that each index $ S_{i_{1},\ldots,i_{s}}$ is a measure of sensitivity describing the portion of the total variance due to the uncertainties in the set of input parameter $\{i_{1},\ldots,i_{s}\}$. With this formula, one can understand which variables are the most and least influential to the output. Furthermore, one can also investigate the interaction between variables.

It is sometimes necessary to rank the total contribution of each variable to the output. For this purpose, the total effect of the input parameter, denoted as the \textit{total sensitivity indices}~\cite{homma1996importance}, can be used. Total sensitivity indices are defined by
\begin{equation}
 S_{i}^{T}= S_{i}+\sum_{j<i}S_{j,i} + \sum_{j<k<i}S_{j,k,i}+\ldots+S_{1,\ldots,n}.
\end{equation}
The total sensitivity indices measure the total contribution of each variable to the output variance which includes all variances caused by its interaction with any other input variable. For example, if $n=3$ we have

\begin{equation}
 S_{1}^{T} = S_{1}+S_{1,2}+S_{1,3}+S_{1,2,3}.
\end{equation}

To compute the value of Sobol indices for a given problem, the most standard method is MCS which offers simplicity and ease of implementation. However, this might come with the high price of computational cost. This is because MCS typically needs more than a thousand simulations to obtain an accurate value of the Sobol indices~\cite{xiu2003modeling}. This is clearly infeasible if the computational time for each deterministic simulation is time demanding. To handle this, PCE can be used as a metamodel to accelerate the computation of the Sobol indices. More specifically, the MF scheme of non-intrusive PCE is considered as a tool to accelerate SA in this paper.

\subsection{Non-Intrusive PCE}

\subsubsection{PCE}
PCE approximates the relationship between the stochastic response output $Y$ and each of the random inputs $\boldsymbol{\xi}$ with the following expansion:
\begin{equation}
Y= f(\boldsymbol{\xi}) = \sum\limits_{j=0}^{\infty} \alpha_{j} \Psi_{j}(\xi),
\end{equation}
where $\alpha$ and $\Psi$ are the PCE coefficients and multivariate polynomials as the product of one-dimensional  orthogonal polynomial, respectively. Consider an index defined by $\boldsymbol{\varphi} = \{\varphi_{1},\ldots,\varphi_{n}\}$ and an index set $\mathcal{I}_{\bold{p}}$, the expansion has be truncated for practical purpose so the expression becomes
\begin{equation}
Y = f(\boldsymbol{\xi}) \cong \sum_{\boldsymbol{\varphi}\in\mathcal{I}_{\bold{p}}} \alpha_{\boldsymbol{\varphi}} \Psi_{\boldsymbol{\varphi}}(\boldsymbol{\xi}).
\end{equation}

The orthogonal polynomial sequence is a family of polynomials such that the collection of polynomials in the sequence are orthogonal to each other by the following relation:
\begin{equation}
\big\langle\Psi_{i}(\boldsymbol{\xi})\Psi_{j}(\boldsymbol{\xi})\big\rangle = \int_{\boldsymbol{\Omega}}\Psi_{i}(\boldsymbol{\xi})\Psi_{j}(\boldsymbol{\xi})\rho(\boldsymbol{\xi})\mathrm{d}\boldsymbol{\xi} = \delta_{ij}, i\neq j,
\end{equation}
where $\delta_{ij}$=1 if $i=j$ and 0 if $i\neq j$. The type of the polynomials used depends on the given probability distribution. Table~\ref{PDFtable} shows the Askey scheme of some continuous hyper-geometric polynomials corresponding to their probability density function~\cite{ askey1985some, xiu2002wiener} . 
\begin{table}
\centering
\begin{tabular}{ccc} \hline 
Distribution type  & Polynomial chaos  & Support \\ \hline
Gaussian & Hermite chaos & $(-\infty,\infty)$      \\ 
Gamma & Laguerre chaos  & [0,$\infty$)       \\ 
Beta & Jacobi chaos & [$a,b$]      \\ 
Uniform & Legendre chaos  & [$a,b$]     \\ \hline 
\end{tabular}
\caption{Standard forms of continuous PDF and their Askey scheme polynomials.}
\label{PDFtable}
\end{table}

A tensor product expansion that includes all combinations of one-dimensional polynomial bases can be utilized in order to extend the PCE to multi-variable approximation. The polynomial bases can then be expanded by using the tensor product operator of order $\bold{p}=\{p_{1},\ldots,p_{n}\}$ via
\begin{equation}
\mathcal{I}_{\bold{p}} \equiv \{\boldsymbol{\varphi}\in  \mathbb{N}^{n}:\varphi_{j}\leq p_{j}, j = 1,\ldots,n\}.
\end{equation}
The total number of polynomial terms $N_{t}$ for the tensor product expansion is calculated by
\begin{equation}
 N_{t}=\prod_{i=1}^{n}(p_{i}+1).
\end{equation}

Total order expansion can be used as an alternative to tensor product operator~\cite{eldred2009comparison}. For the total-order expansion of order $p$, the index set is defined by
\begin{equation}
\mathcal{I}_{p} \equiv \{\boldsymbol{\varphi}\in  \mathbb{N}^{n}:|\boldsymbol{\varphi}|\leq p\},
\end{equation}
where $|\boldsymbol{\varphi}|=\varphi_{1}+\ldots+\varphi_{n}$. The total number of polynomial terms $N_{t}$ for the total order expansion is computed by
\begin{equation}
N_{t} = \frac{(n+p)!}{n!p!}.
\end{equation}

After the PCE has been built, the next task is to compute the coefficients.

\subsubsection{Spectral-projection based PCE}
In this paper, we focus on the spectral projection method that calculates the coefficients based on the orthogonality of the bases. The coefficients are computed via

\begin{equation}
\alpha_{i} = \frac{\big \langle f(\boldsymbol{\xi}),\Psi_{i}(\boldsymbol{\xi}) \big \rangle}{\big \langle\Psi_{i}^{2}(\boldsymbol{\xi})\big \rangle} = \frac{1}{\big \langle\Psi_{i}^{2}(\boldsymbol{\xi})\big \rangle}\int_{\boldsymbol{\Omega}}f(\boldsymbol{\xi})\Psi_{i}(\boldsymbol{\xi})\rho(\boldsymbol{\xi})\mathrm{d}\boldsymbol{\xi}.
\end{equation}

The main part of the work is the calculation of multi-dimensional integral in the numerator. For this purpose, tensor-product quadrature can be efficiently employed to perform the integration in low dimensionality problems ($n\leq3$). However, tensor-product quadrature might produce a large number of collocation points in problems with high-dimensionality. Sparse grid quadrature is an important alternative to tensor product quadrature in performing this integration. In this paper, we use sparse grid integration to compute the numerator in all examples. By using a sparse grid rule, the number of collocation points is reduced if compared to the tensor-product rule, thus, making it feasible to use spectral-projection based PCE in high-dimension. Moreover, if a sparse grid is employed, it is necessary to use the polynomial bases built from the sum of tensor product expansion to avoid numerical noise in the coefficients of higher order terms~\cite{constantine2012sparse}; we used this approach in all examples. 

We define the isotropic sparse grid  at level $c$ where $c=0,1,2,\ldots$ ~\cite{smolyak1960interpolation, bungartz2004sparse, garcke2012sparse, griebel1990combination} by
\begin{equation}
 \mathcal{A}_{c,n}(f) = \sum_{c-n+1\leq |\boldsymbol{\varphi}|\leq c} (-1)^{c-|\boldsymbol{\varphi}|}
 \bigg(
 \begin{array}{c}
  n-1\\
  c-|\boldsymbol{\varphi}|
 \end{array}
 \bigg)
 \boldsymbol{\mathcal{Q}}_{\boldsymbol{\varphi}}(f),
\end{equation}
where $\boldsymbol{\mathcal{Q}_{\varphi}}$ are the tensor product quadrature formulas for sparse grid equation. The form of formulas $\boldsymbol{\mathcal{Q}_{\varphi}}$ depends on the type of polynomial used in PCE. For example, Gauss-Legendre and Gauss-Hermite quadrature are used when Legendre and Hermite polynomials are employed in the PCE approximation, respectively.

Another way to express sparse grid is in terms of difference formulas~\cite{gerstner1998numerical}
\begin{equation}
 \mathcal{A}_{c,n}(f) = \sum_{|\boldsymbol{\varphi}|\leq c}\boldsymbol{\Delta}_{\boldsymbol{\varphi}}(f),
\end{equation}
where $\boldsymbol{\Delta}_{\boldsymbol{\varphi}}=(\Delta_{\varphi_{1}}\otimes\ldots\otimes\Delta_{\varphi_{n}})(f)$ and $\Delta_{\varphi_{k}}=\mathcal{Q}_{\varphi_{k}}-\mathcal{Q}_{\varphi_{k-1}} $ with $\mathcal{Q}_{-1}=0$.

\subsubsection{Multi-fidelity extension}
The PCE can be extended into the MF version by employing a combination of the LF and correction PCE~\cite{ng2012multifidelity}. The correction PCE, which serves as a corrector for LF-PCE, is built by using the difference between the LF and HF function at selected sample locations. The combination can be in the form of additive or multiplicative form~\cite{ng2012multifidelity}. In this work, we only use the additive form defined by
\begin{equation}
CR(\boldsymbol{\xi}) = f_{\text{high}}(\boldsymbol{\xi})-f_{\text{\text{low}}}(\boldsymbol{\xi}),
\end{equation}
so
\begin{equation}
f_{\text{high}}(\boldsymbol{\xi}) = f_{\text{\text{low}}}(\boldsymbol{\xi}) + CR(\boldsymbol{\xi}), 
\end{equation}
where $CR(\boldsymbol{\xi})$ is the difference between the LF ($f_{\text{\text{low}}}$) and HF function ($f_{\text{high}}$), which serves as a correction term. 

In integrating sparse grid with PCE, we first define $S_{w,n}[f]$ as the PCE of $f(\boldsymbol{\xi})$ at sparse grid level $w$ with dimension $n$. Within MF framework, the HF-PCE $S_{w,n}[f_{\text{high}}]$ can be constructed by 
\begin{equation}
 S_{w,n}[f_{\text{high}}] = S_{w,n}[f_{\text{low}}]+S_{w-q,n}[CR],
\end{equation}
where $S_{w,n}[f_{\text{low}}]$ and $S_{w-q,n}[CR]$ are the LF and correction PCE, respectively. Here, $q<w$ is defined as sparse grid level offset between the LF and correction PCE.

If $\alpha_{\text{\text{low}}}$ and $\alpha_{CR}$ are the coefficients of the LF and correction PCE, respectively, and $\Gamma_{w,n}$ is the set of multi-indices of the $n-$dimensional PCE on sparse grid level $w$, we can then formulate the following:

\begin{equation}
	S_{w,n}[f_{\text{low}}](\boldsymbol{\xi}) = \sum_{\boldsymbol{\varphi}\in\Gamma_{w,n}}\alpha_{\text{low}_{\boldsymbol{\varphi}}}\Psi_{\boldsymbol{\varphi}}(\boldsymbol{\xi}),
\end{equation}

\begin{equation}
S_{w-q,n}[CR](\boldsymbol{\xi}) = \sum_{\boldsymbol{\varphi}\in\Gamma_{w-q,n}}\alpha_{CR_{\boldsymbol{\varphi}}}\Psi_{\boldsymbol{\varphi}}(\boldsymbol{\xi}).
\end{equation}
The MF polynomial expansion can then be expressed as
\begin{equation}
\label{eq:mfpce}
S_{w,n}[f_{\text{low}}](\boldsymbol{\xi})+S_{w-q,n}[CR](\boldsymbol{\xi}) = \sum_{\boldsymbol{\varphi}\in\Gamma_{w-q,n} } (\alpha_{\text{\text{low}}_{\boldsymbol{\varphi}}} + \alpha_{CR_{\boldsymbol{\varphi}}})\Psi_{\boldsymbol{\varphi}}(\boldsymbol{\xi}) + \sum_{ \boldsymbol{\varphi}\in\Gamma_{w,n} \backslash \Gamma_{w-q,n}} (\alpha_{\text{\text{low}}_{\boldsymbol{\varphi}}})\Psi_{\boldsymbol{\varphi}}(\boldsymbol{\xi}),
\end{equation}
where \(\Gamma_{w-q,n} \subset \Gamma_{w,n}\) are the common bases of the LF and correction expansion. 

\begin{figure}[H]
\centering
\begin{subfigure}{.33\columnwidth}
\includegraphics[width=1\columnwidth]{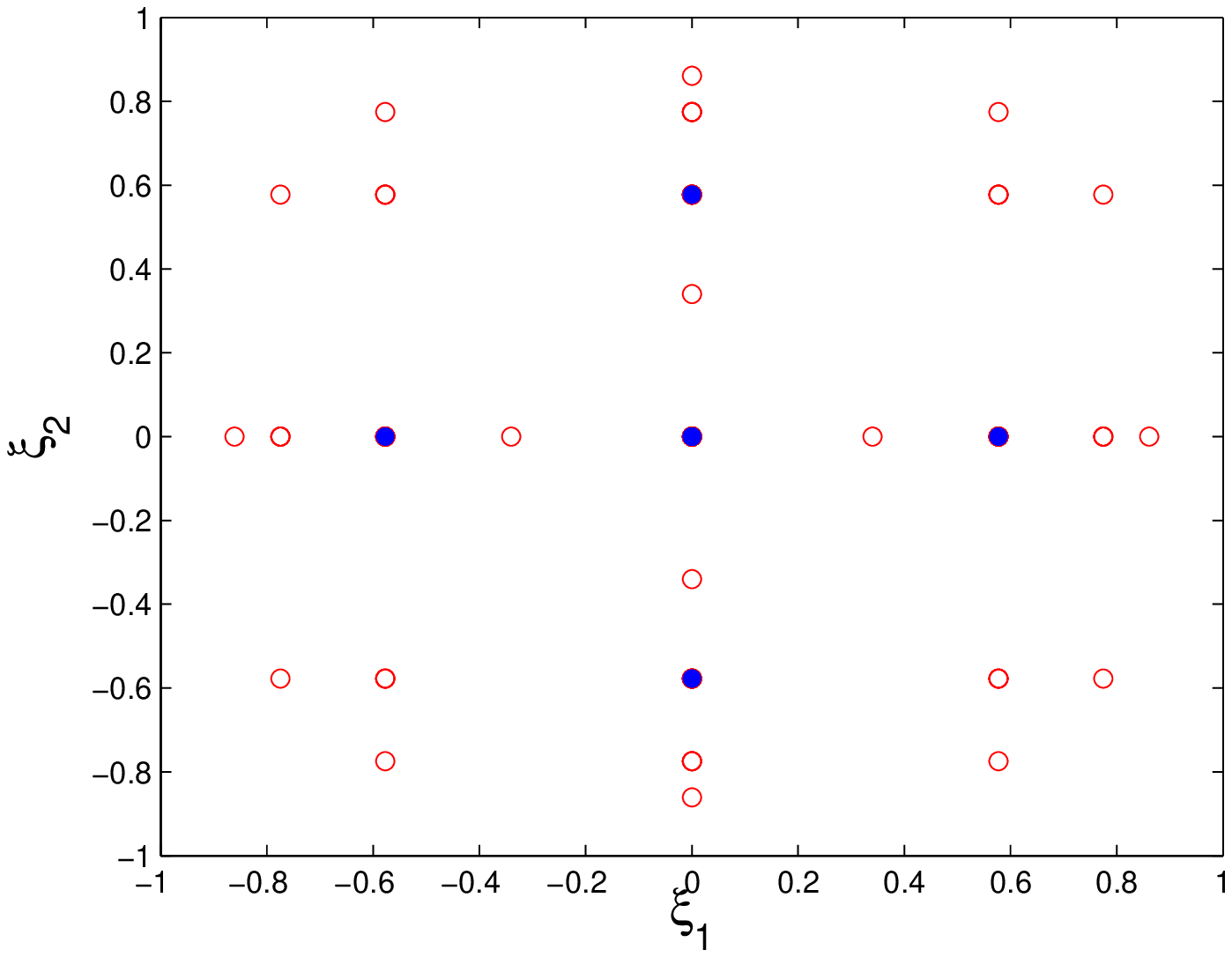}%
\caption{$w=3,q=2$.}%
\label{fig:w3q1}%
\end{subfigure}\hfill%
\begin{subfigure}{.33\columnwidth}
\includegraphics[width=1\columnwidth]{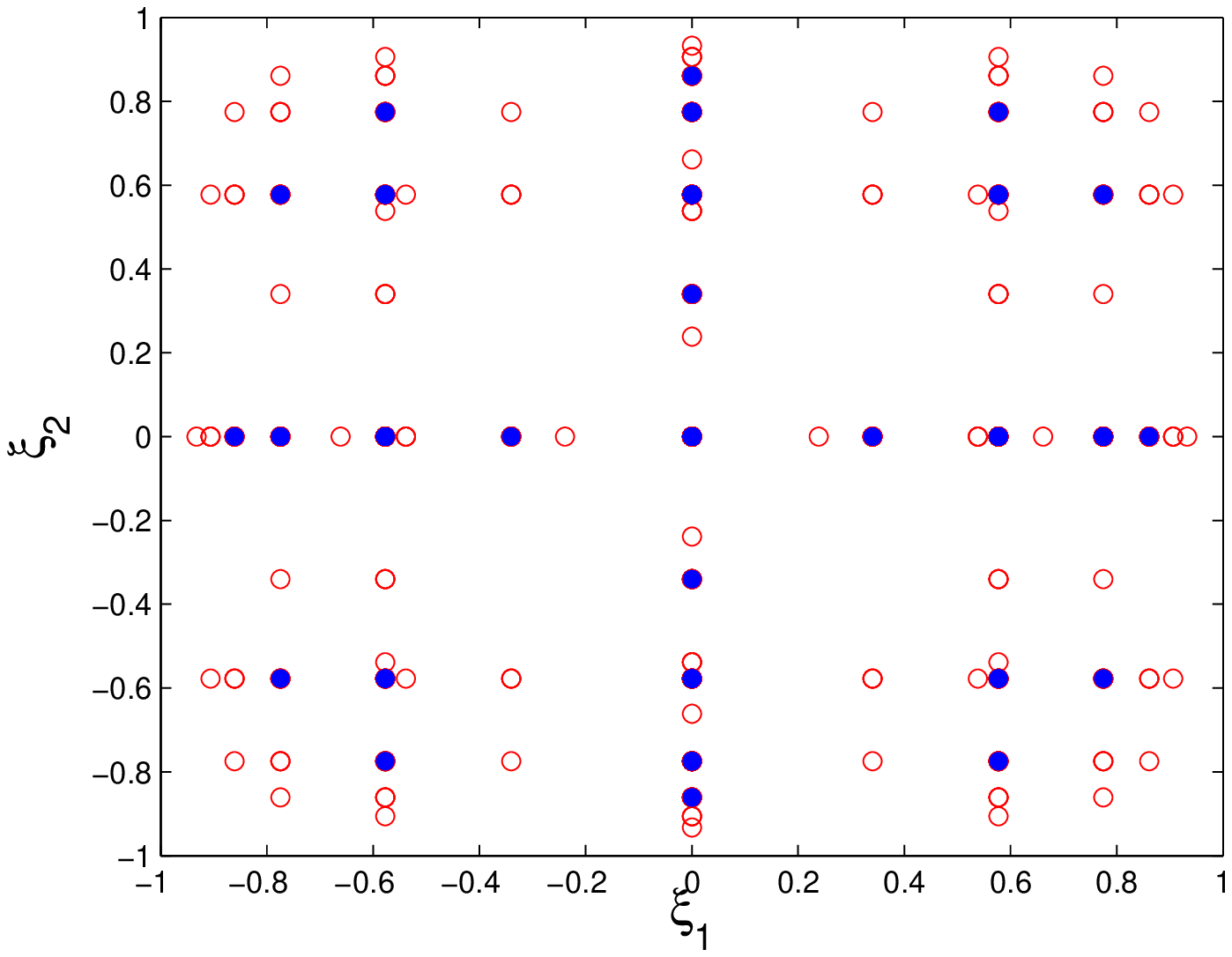}%
\caption{$w=5,q=2$.}%
\label{fig:w5q2}%
\end{subfigure}\hfill%
\begin{subfigure}{.33\columnwidth}
\includegraphics[width=1\columnwidth]{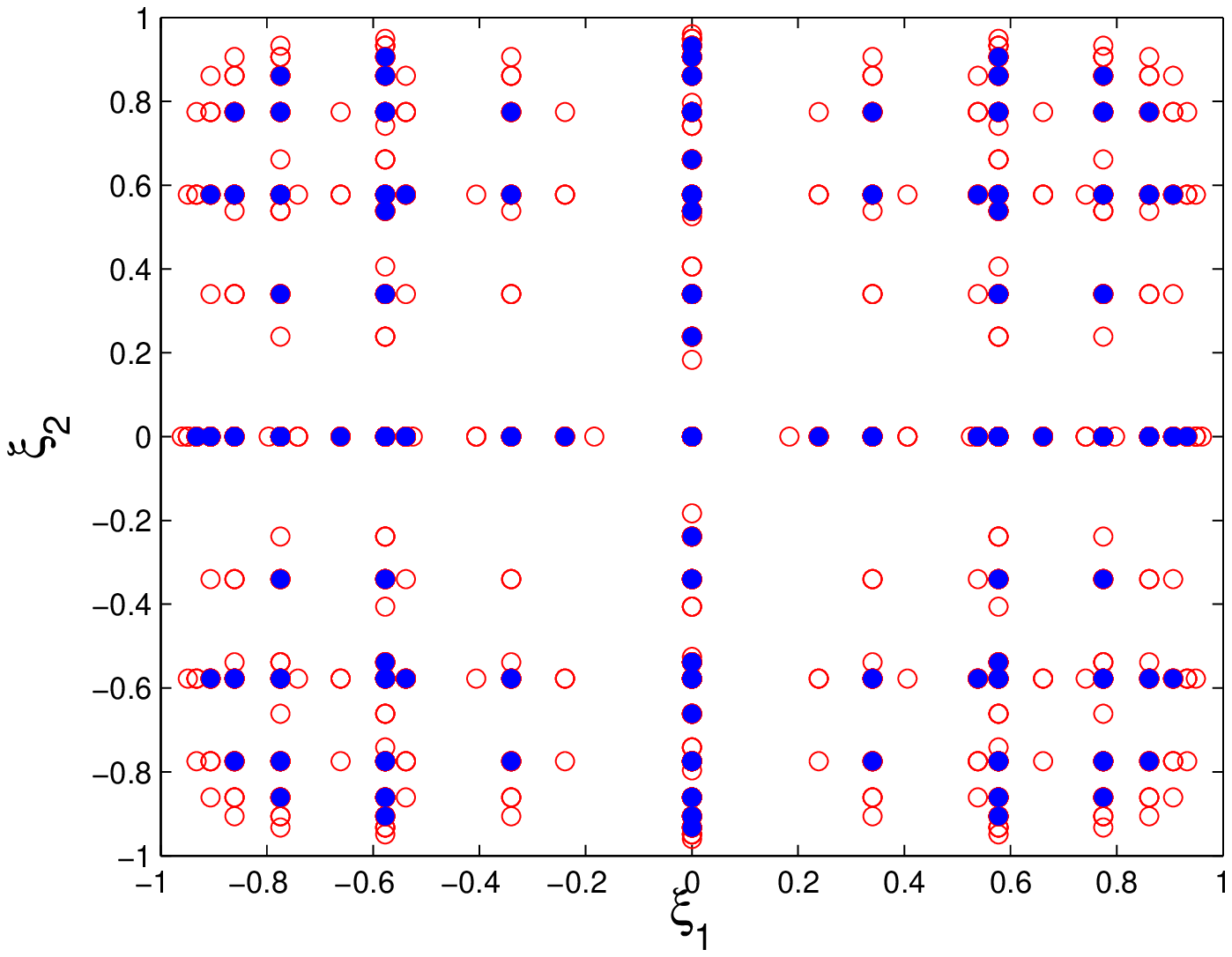}%
\caption{$w=7,q=2$.}%
\label{fig:w7q2}%
\end{subfigure}\hfill%
\caption{Examples of sparse grid for  MF-PCE where blue filled and red circles are HF and LF samples, respectively. The HF sparse grid is always on a lower sparse grid level than that of the LF one. }
\label{fig:mfspex}
\end{figure}

After the PCE has been built and the coefficients been computed, the statistical moments can then be directly computed from the PCE coefficients. The sparse grid growth rule used in this paper is $2m+1$, where $m \geq 0$.
Figure~\ref{fig:mfspex} shows various examples of the sparse grids with various $w$ values and $q=2$ for MF-PCE purpose. It can be seen in this figure that the HF samples (lower level sparse grid) are the subset of the LF samples (higher level sparse grid), thus made it possible to correct the LF values in the intersecting samples.

\subsubsection{Calculation of statistical moments}
The mean for the standard single HF-PCE is simply the first PCE coefficient which is given by
\begin{equation}\label{eq:2.21}
\mu_{f}=E[f(\boldsymbol{\xi})] = \int_{\boldsymbol{\Omega}}f(\boldsymbol{\xi})\rho(\boldsymbol{\xi})\mathrm{d}\boldsymbol{\xi} \approx \alpha_{0}.
\end{equation}
The variance can be obtained as follows: 
\begin{equation}\label{eq:2.22}
\sigma_{f}^{2} = \mathbb{V}[(f(\boldsymbol{\xi}))] = \int_{\boldsymbol{\Omega}}(f(\boldsymbol{\xi})-\mu_{f})^2\rho(\boldsymbol{\xi})\mathrm{d}\boldsymbol{\xi} \approx \sum_{j=1}^{P} \alpha_{{j}}^2 \langle \Psi_{j}^{2}(\boldsymbol{\xi}) \rangle = D_{PC},
\end{equation}
where $P+1$ is the size of polynomial basis. 

For the MF-PCE expression shown in Eq.~\ref{eq:mfpce}, the moments are calculated via the following formulations:
\begin{equation}\label{eq:mfm}
\mu_f \approx \alpha_{\text{\text{low}}_{0}} + \alpha_{CR_{0}},
\end{equation}
\begin{equation}\label{eq:mfs}
\sigma_{f}^{2} \approx \sum_{\boldsymbol{\varphi}\in\Gamma_{w-q,n} \backslash 0} (\alpha_{\text{\text{low}}_{\boldsymbol{\varphi}}} + \alpha_{CR_{\boldsymbol{\varphi}}})^{2} \langle \Psi_{\boldsymbol{\varphi}}^{2}(\boldsymbol{\xi}) \rangle + \sum_{\boldsymbol{\varphi}\in\Gamma_{w,n}\backslash\Gamma_{w-q,n}} (\alpha_{\text{\text{low}}_{\boldsymbol{\varphi}}})^{2} \langle \Psi_{\boldsymbol{\varphi}}^{2}(\boldsymbol{\xi}) \rangle = D_{PC}.
\end{equation}
Furthermore, as the core of this work, the PCE coefficients are used in the computation of the PCE-based Sobol indices.
\subsubsection{PCE-based Sobol indices}
Obtaining the PCE-based Sobol indices is straightforward when the PCE coefficients for a given expansion are already obtained~\cite{sudret2008global}. This can be done by firstly deriving the Sobol decomposition of the PCE approximation of $f(\boldsymbol{\xi})$. First, we define $\mathcal{L}_{i_{1},\ldots,i_{s}}$ as the set of $\boldsymbol{\gamma}$ tuples, with the indices $(i_{1},\ldots,i_{s})$ are nonzero, as 
\begin{equation}
 \mathcal{L}_{i_{1},\ldots,i_{s}} = \Bigg\{  \boldsymbol{\gamma} :
 \begin{aligned}
  \gamma_{k}>0  ~~ & \forall k=1,\ldots,n, & k \in (i_{1},\ldots,i_{s}) \\
  \gamma_{j}=0  ~~ & \forall k=1,\ldots,n, & k \cancel{\in} (i_{1},\ldots,i_{s}) \\
 \end{aligned}
 \Bigg\}.
\end{equation}
By gathering the $P$ polynomials according to the parameters they depend on, the Sobol decomposition of PCE can then be reads as
\begin{equation}
\begin{aligned}
 f_{PC}(\boldsymbol{\xi}) = \alpha_{0} +\sum_{i=1}^{n}\sum_{\boldsymbol{\gamma}\in\mathcal{L}_{i}}\alpha_{\boldsymbol{\gamma}}\Psi_{\boldsymbol{\gamma}}(\xi_{i}) + \sum_{1\leq i_{1}<i_{2} \leq n} \sum_{\boldsymbol{\gamma}\in \mathcal{L}_{i_{1},i_{2}}} \alpha_{\boldsymbol{\gamma}}\Psi_{\boldsymbol{\gamma}}(\xi_{i_{1}},\xi_{i_{2}})+\ldots  \\
 +\sum_{1\leq i_{1}<\ldots<i_{s} \leq n} \sum_{\boldsymbol{\gamma}\in \mathcal{L}_{i_{1},\ldots,i_{s}}} \alpha_{\boldsymbol{\gamma}}\Psi_{\boldsymbol{\gamma}}(\xi_{i_{1}},\ldots,\xi_{i_{s}})\\ 
  +\ldots+\sum_{\boldsymbol{\gamma} \in \mathcal{L}_{1,2,\ldots,n}} \alpha_{\boldsymbol{\gamma}}\Psi_{\boldsymbol{\gamma}}(\xi_{1},\xi_{n}).
 \end{aligned}
\end{equation}
Based on the above decomposition, the PCE-based Sobol indices can then be simply defined as
\begin{equation}
 SU_{i_{1},\ldots,i_{s}} = \sum_{\boldsymbol{\gamma} \in \mathcal{L}_{i_{1},\ldots,i_{s}}} \alpha_{\boldsymbol{\gamma}}^{2}E[\Psi_{\boldsymbol{\gamma}}^{2}]/D_{PC}.
\end{equation}

It is clear from the above equation that the PCE-based Sobol indices can be directly obtained when the PCE coefficients are available. Using a similar procedure, we can then easily derive the expression for PCE-based total sensitivity indices. The PCE-based total sensitivity indices then can be reads as
\begin{equation}
 SU_{i}^{T}= SU_{i}+\sum_{j<i}SU_{j,i}+ \sum_{j<k<i}SU_{j,k,i}+\ldots+SU_{1,\ldots,n},
\end{equation}
which can be simplified into
\begin{equation}
 SU_{i}^{T}=\frac{1}{D_{PC}}\sum_{\boldsymbol{\gamma}\in\mathcal{I}_{i}^{+}}\alpha_{\boldsymbol{\gamma}}^{2},
\end{equation}
where $\mathcal{I}_{i}^{+}$ is the set of all indices with a non-zero $i-$th components, formally defined as
\begin{equation}
 \mathcal{I}_{i}^{+}\equiv \{\boldsymbol{\gamma}\in\mathbb{N}^{n}:0\leq|\boldsymbol{\gamma}|\leq p,\gamma_{i}\neq 0\}.
\end{equation}

In MF-PCE, the polynomial terms and coefficients involved in Sobol indices computation are built from the LF and the correction expansion shown in Eq.~\ref{eq:mfpce}. The process starts by building the sparse grid and the corresponding polynomial bases for LF-PCE followed by correction PCE. After the two PCEs are combined into a single MF-PCE, post processing can be directly performed to calculate the statistical moments and the Sobol indices.

\section{Results}
\label{sec:3}
The capability of MF-PCE on algebraic cases when the ``high'' and ``low'' fidelity models are available are studied. In the algebraic case considered, the term of ``high'' and ``low'' fidelity models does not mean that the models are expensive and cheap in the real sense. The LF model is artificially constructed so it represents the HF model but with a discrepancy, which is measured by the $\text{r}_{lh}^2$ correlation value and mean absolute relative error (MARE) between the LF and HF function $\text{MARE}_{lh}$.  These two statistical measures can be used to perform a preliminary analysis of whether the LF function is similar to the HF one or not~\cite{toal2015some}. The equation for $\text{r}_{lh}^2$ correlation and $\text{MARE}_{lh}$ are
\begin{equation}
 \text{r}_{lh}^{2} = \bigg(\frac{\sum_{i=1}^{N_{v}}(y_{h_{i}}-\bar{y}_{h})({y_{l_{i}}-\bar{y}_{l}})}{\sqrt{\sum_{i=1}^{N_{v}}(y_{h_{i}}-\bar{y}_{h})^{2}} \sqrt{\sum_{i=1}^{N_{v}}(y_{l_{i}}-\bar{y}_{l})^{2}}} \bigg)^{2}
\end{equation}
and 
\begin{equation}
 \text{MARE}_{lh} = \frac{1}{N_{v}} \sum_{i=1}^{N_{v}}\bigg|\frac{y_{l_{i}}-{y_{h_{i}}}}{y_{h_{i}}}\bigg|,
\end{equation}
respectively, where $y_{h}$ and $y_{l}$ are a set of $N_{v}$ observations of the HF and LF data for identical inputs with the bar ($\bar{y}_{h}, \bar{y}_{l}$) denoting the mean of these sets.

Analysis of the decay of PCE coefficients is a proper way to analyze whether the LF function can improve the capability of MF-PCE in approximating the HF function~\cite{ng2012multifidelity}. The LF function can assist MF-PCE if the decay of the correction PCE coefficients is faster than that of the HF one~\cite{ng2012multifidelity}. This indicates that the correction function is less complex than the HF function. Basically, this means that for the same level of sparse grid, the combination of correction and LF-PCE captures more relevant information regarding the HF function than HF-PCE. As an example, consider a case where the HF and LF function is perfectly correlated with $r_{lh}^{2}$ equals to one with an offset in the magnitude. This basically means that the correction function is just a constant which can be approximated by simply a zeroth order polynomial. Assuming that the LF function is much cheaper than the HF function so we can construct a high-level sparse grid for LF-PCE, one can build a highly accurate MF-PCE by combining this LF-PCE with this constant correction function. In this paper, we plot the coefficients of correction, LF-, and HF-PCE to do this analysis. To analyze the decay of the PCE coefficients, the absolute value of the coefficients must be sorted first and then plotted in a single plot.

We measured the error of total and all Sobol sensitivity indices, expressed as
\begin{equation}
\text{e} = \sum_{i=1}^{n_{si}} |SU^{(i)}-\hat{SU}^{(i)}|
\end{equation}
and
\begin{equation}
 \text{e}_{\text{T}} = \sum_{i=1}^{n} |SU_{i}^{T}-\hat{SU}_{i}^{T}|,
\end{equation}
respectively, where $n_{si}=2^{n}-1$ is the number of Sobol indices for a given dimension. Here, $\hat{SU}$ and $\hat{SU}^{T}$ are the reference value for the single and total Sobol indices, respectively. For implementation, the value of $\hat{SU}$ and $\hat{SU}^{T}$ are determined by using analytical calculation (such as in the Ishigami function) or PCE with a sparse grid of sufficiently high level.

Moreover, in order to assess the quality of the PCE approximation, the $\text{r}^{2}$ and MARE of the prediction to the true function is used, which are defined by
\begin{equation}
 \text{r}^{2} = \bigg(\frac{\sum_{i=1}^{N_{v}}(y_{h_{i}}-\bar{y}_{h})({\hat{y}_{h_{i}}-\bar{\hat{y}}_{h}})}{\sqrt{\sum_{i=1}^{N_{v}}(y_{h_{i}}-\bar{y}_{h})^{2}} \sqrt{\sum_{i=1}^{N_{v}}(\hat{y}_{h_{i}}-\bar{\hat{y}}_{h})^{2}}} \bigg)^{2}
\end{equation}
and
\begin{equation}
\text{MARE} = \frac{1}{N_{v}} \sum_{i=1}^{N_{v}}\bigg|\frac{\hat{y}_{h_{i}}-{y_{h_{i}}}}{y_{h_{i}}}\bigg|,
\end{equation}
respectively, where $\hat{y}_{h_{i}}$ is the output of the PCE prediction. Notice that $\text{r}^{2}$ and MARE without any subscript denote the prediction error, and not the discrepancy between the LF and HF function.

In the following examples, the sparse grid level for HF-PCE is written as ``SG-$w$'', while for MF-PCE it is written as ``SG-$(w$-$q$)-$w$''. The convergence of the errors is plotted with the $x-$axis as the number of function evaluations $n_{e}$. The calculation of $n_{e}$ does not distinguish the HF or LF function evaluation when HF- and LF-PCE are considered since it does not make sense to do so in algebraic problems. However, when calculating $n_{e}$ for MF-PCE in algebraic function, the cost of LF function is not considered to make a fair comparison between both the HF- and MF-PCE from the algorithmic point of view. Furthermore, we also compare the performance of HF- and MF-PCE in estimating  Sobol indices with the hypothetical cost of the LF simulation is taken into account. This hypothetical cost of the LF simulation is evaluated in terms of the ratio between the LF and HF evaluation cost (denoted as RT). The purpose is to evaluate the capability and usefulness of MF-PCE in accelerating the computation of Sobol indices when it is applied in a real-world setting, where the cost of the LF simulation is usually not negligible. Here, we use four values for the hypothetical cost, that is, RT=$\frac{1}{4},\frac{1}{8},\frac{1}{16}$ and $\frac{1}{32}$. The total simulation cost of MF-PCE (i.e., $n_{tot}$) is then the sum of the HF and LF evaluation cost, in the unit equivalent to one HF evaluation function. 

\subsection{Borehole function}
The first example is the borehole function, which is an eight-dimensional function that models water flow through a borehole~\cite{morris1993bayesian}. Mathematically it is expressed as

\begin{equation}\label{eq:4.10}
f_{h}(\boldsymbol{\xi}) = \frac{2\pi T_{u}(H_{u}-H_{l})}{\text{ln}(r_{a}/r_{w})\left(1+\frac{2LT_{u}}{\text{ln}(r_{a}/r_{w})r_{w}^{2}K_{w}}+\frac{T_{u}}{T_{l}}\right)},
\end{equation}
where the various parameters are as defined in Table \ref{tbl:borehole}.

The LF borehole function~\cite{xiong2013sequential} is expressed as
\begin{equation}\label{eq:4.11}
f_{l}(\boldsymbol{\xi}) = \frac{5 T_{u}(H_{u}-H_{l})}{\text{ln}(r_{a}/r_{w})\left(1.5+\frac{2LT_{u}}{\text{ln}(r_{a}/r_{w})r_{w}^{2}K_{w}}+\frac{T_{u}}{T_{l}}\right)}.
\end{equation}

For SA purpose, the random variables are all uniformly distributed as shown in Table \ref{tbl:borehole}.
\begin{table}[h]
\centering
\begin{tabular}{ccc} \hline
Random variable & Definition & Probability distribution \\ \hline 
$r_{w}$ & radius of borehole (m)  & Uniform [0.05, 0.15]   \\ 
$r_{a}$ & radius of influence (m) &  Uniform [100, 50000]  \\ 
$T_{u}$ & transmissivity of upper aquifer (m\textsuperscript{2}/yr) & Uniform [63700, 115600]  \\ 
$H_{u}$ & potentiometric head of upper aquifer (m) &  Uniform [990, 1100]  \\ 
$T_{l}$ & transmissivity of lower aquifer (m\textsuperscript{2}/yr)  & Uniform [63.1, 116]   \\ 
$H_{l}$ & potentiometric head of lower aquifer (m)  & Uniform [700, 820]  \\ 
$L$ &  length of borehole (m) & Uniform [1120, 1680]  \\ 
$K_{w}$ & hydraulic conductivity of borehole (m/yr)  & Uniform [9855, 12045]   \\ \hline
\end{tabular}
\caption{Random variable distributions for the borehole test case.}
\label{tbl:borehole}
\end{table}

The LF Borehole function exhibits an almost perfect correlation with the HF one (r$_{lh}^{2}=0.999$), which means that the trend response of the LF function is similar to the HF one, but with response value offset (MARE$_{lh}$=0.204). To acquire monotonous convergence of the Sobol indices error, the reference values for errors calculation are obtained using level 5 HF sparse grid, in which the first order and total sensitivity indices computed using this high-level sparse grid are shown in Table~\ref{tbl:boreholesobol} (note that the analytical Sobol indices are not available for the borehole function). Since the total Sobol indices are not equal to the first order, this means that the borehole function exhibits a correlation between the variables that need to be captured by sparse grid level higher than one. The errors of Sobol sensitivity indices are then computed using these reference values. For this test function, the sparse grid level offset $q$ is set to 1.

\begin{table}[H]
\renewcommand{\arraystretch}{1.3}
\caption{Reference Sobol indices for the borehole function obtained using HF-PCE SG-5 for computing the error.}
\label{tbl:boreholesobol}
\centering
\begin{tabular}{cc|cc} \hline
 \multicolumn{2}{c}{Total} &  \multicolumn{2}{c}{First order}\\ \hline
$\hat{SU}_{1}^{T}$ & 0.8668 & $\hat{SU}_{1}$ &  0.8289  \\
$\hat{SU}_{2}^{T}$ & 0 & $\hat{SU}_{2}$ & 0 \\
$\hat{SU}_{3}^{T}$ & 0 & $\hat{SU}_{3}$ & 0  \\
$\hat{SU}_{4}^{T}$ & 0.0541 & $\hat{SU}_{4}$ & 0.0414  \\
$\hat{SU}_{5}^{T}$ & 0 & $\hat{SU}_{5}$ & 0  \\
$\hat{SU}_{6}^{T}$ & 0.0541 & $\hat{SU}_{6}$ & 0.0414  \\
$\hat{SU}_{7}^{T}$ & 0.0521 & $\hat{SU}_{7}$ & 0.0393  \\
$\hat{SU}_{8}^{T}$ & 0.0127 & $\hat{SU}_{8}$ & 0.0095  \\ \hline
\end{tabular}
\end{table}

\begin{figure}[H]
\centering
\begin{subfigure}{.48\columnwidth}
\includegraphics[width=1\columnwidth]{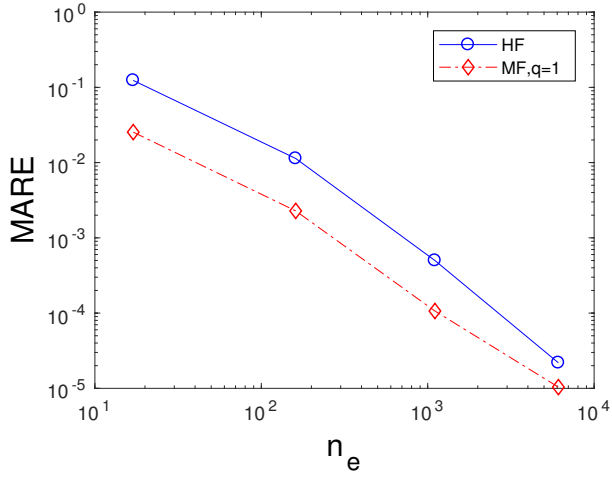}%
\caption{MARE.}%
\label{fig:BOREHOLE_error}%
\end{subfigure}\hfill%
\begin{subfigure}{.48\columnwidth}
\includegraphics[width=1\columnwidth]{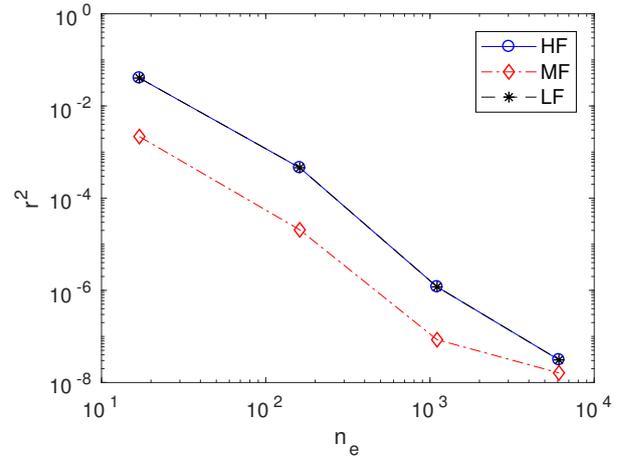}%
\caption{r\textsuperscript{2}.}%
\label{fig:BOREHOLE_R2_error}%
\end{subfigure}\hfill%
\begin{subfigure}{.48\columnwidth}
\includegraphics[width=1\columnwidth]{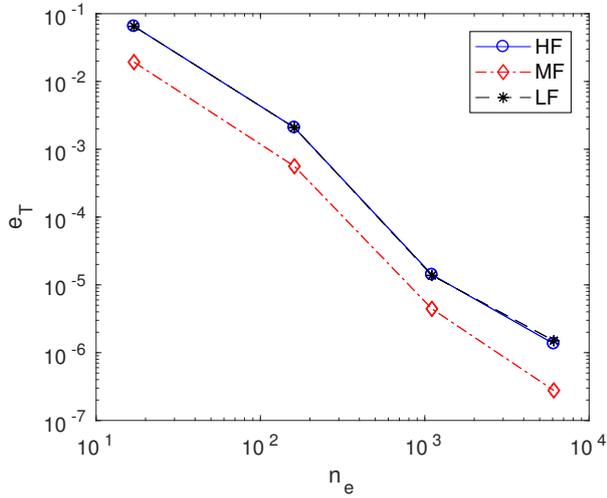}%
\caption{e$_{\text{T}}$.}%
\label{fig:BOREHOLE_T_error}%
\end{subfigure}\hfill%
\begin{subfigure}{.48\columnwidth}
\includegraphics[width=1\columnwidth]{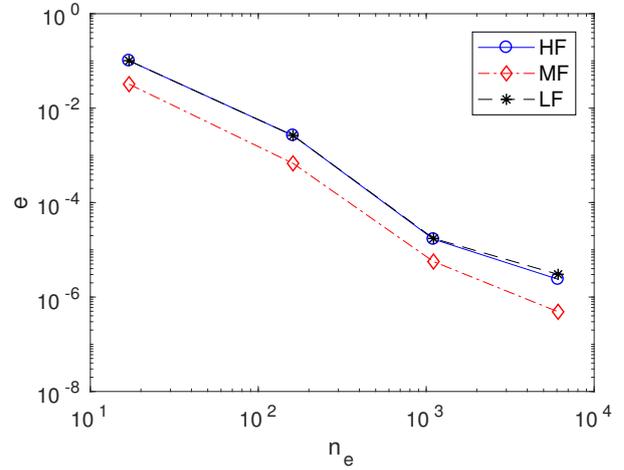}%
\caption{e.}%
\label{fig:BOREHOLE_AS_error}%
\end{subfigure}\hfill%
\caption{Convergence of error metrics for the borehole function with respect to the number of function evaluations.  The MF-PCE method estimates the Sobol indices better than HF-PCE with the same amount of function evaluations, while the estimated values from HF- and LF-PCE are similar to each other.}
\label{fig:boreholeerrorresult}
\end{figure}

The errors of the Sobol indices obtained from various schemes of PCE are shown in Fig.~\ref{fig:boreholeerrorresult}. It can be seen in Figs.~\ref{fig:BOREHOLE_error} and~\ref{fig:BOREHOLE_R2_error} that the MF-PCE method is able to reduce the MARE and r$^{2}$ with the same amount of HF function evaluations with HF-PCE. Results also show that MF-PCE reduces the e and e$_{\text{T}}$ compared to the HF-PCE method with a monotonous decreasing trend. The convergence trends for both e and e$_{\text{T}}$ are roughly similar for all schemes. Nonetheless, the e and e$_{\text{T}}$ obtained from LF-PCE without correction are already very accurate and a good representation of the HF borehole function from the Sobol indices point of view. This means that the trend of the HF borehole function is obviously captured by the LF borehole function, due to the very high correlation value between the LF and HF function. In fact, the convergence of r$^{2}$ is very similar for both HF-PCE and LF-PCE. This suggests that the LF function is adequate to be used for estimating the Sobol indices of the HF function without any correction term if the correlation is near one. However, although the Sobol indices obtained from LF-PCE are a highly accurate representation of the HF function, the obtained statistical moments are inaccurate because the response is not corrected. This means that if the PCE model is to be used for estimating the statistical moments, the LF function alone is not enough and needs correction to be properly used for both UQ and SA purposes (this is why the MARE for LF-PCE is not depicted since it is already obvious).

\begin{figure}[H]
\centering
\begin{subfigure}{.48\columnwidth}
\includegraphics[width=1\columnwidth]{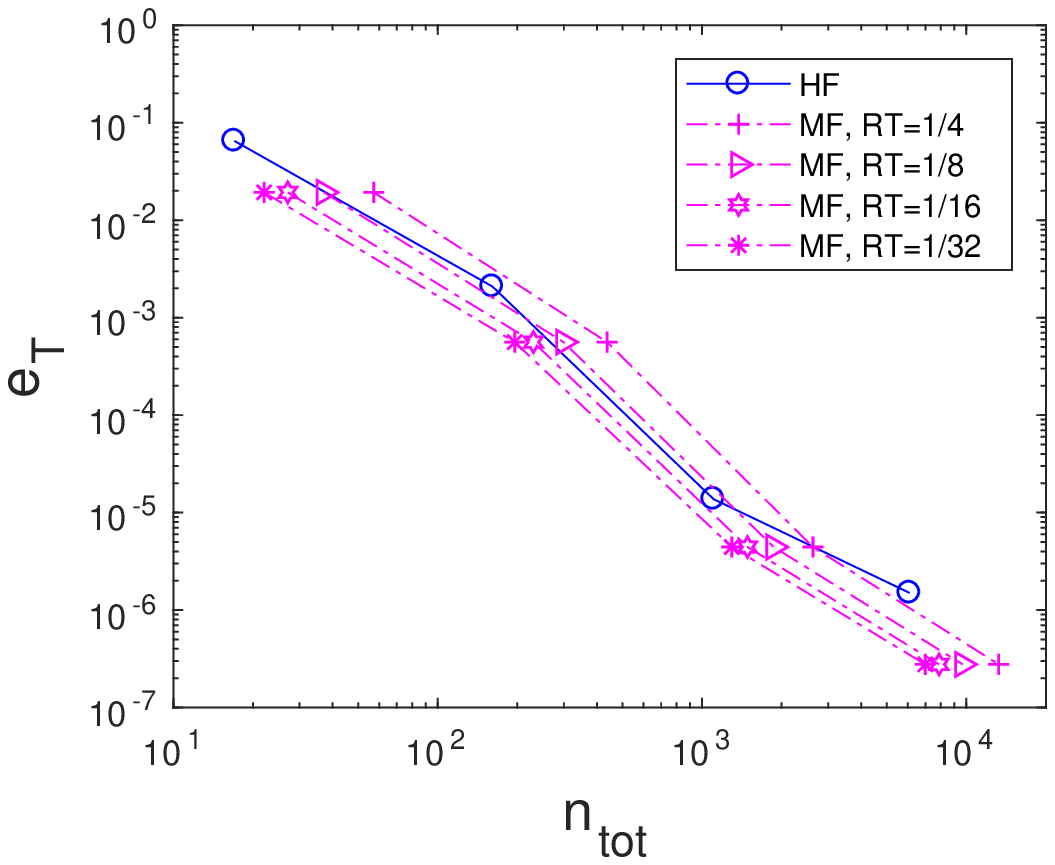}%
\caption{e$_{\text{T}}$.}%
\label{fig:BOREHOLE_TS_error_ntot}%
\end{subfigure}\hfill%
\begin{subfigure}{.48\columnwidth}
\includegraphics[width=1\columnwidth]{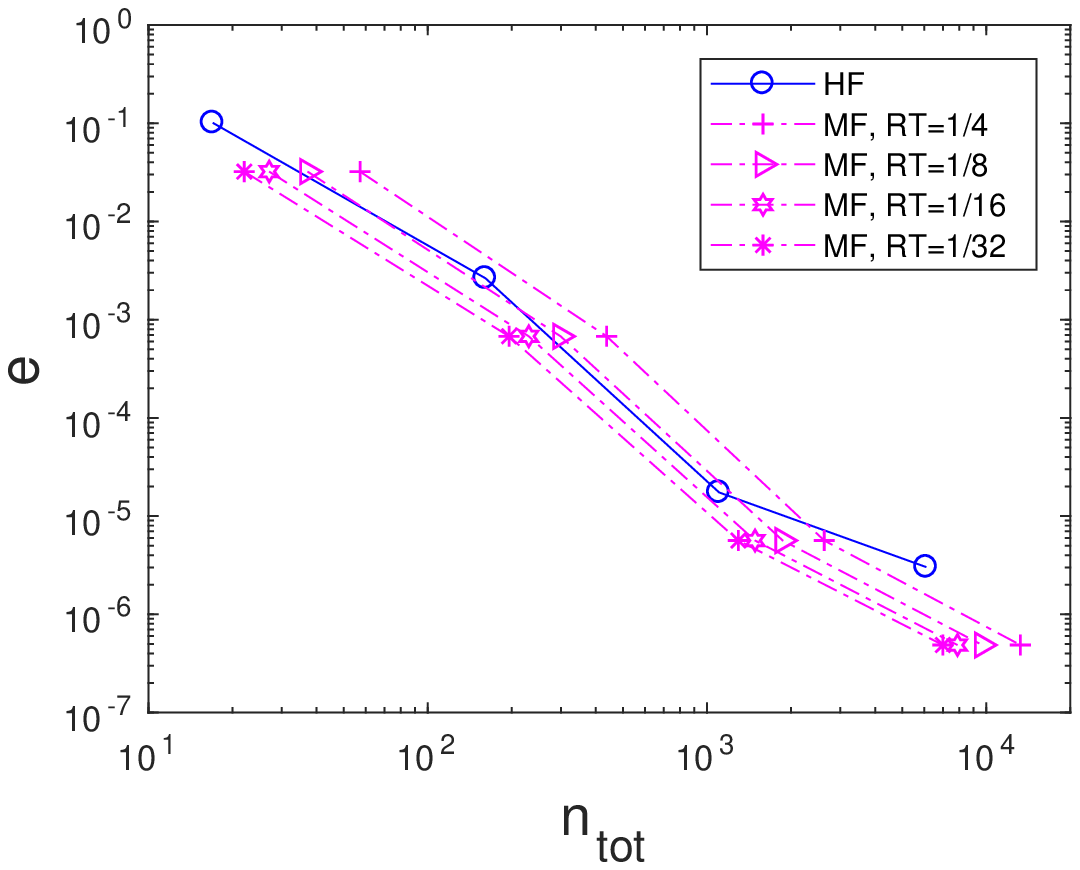}%
\caption{e.}%
\label{fig:BOREHOLE_AS_error_ntot}%
\end{subfigure}\hfill%
\caption{Convergence of error metrics for the borehole function with respect to the total simulation cost assuming a hypothetical LF model cost.}
\label{fig:boreholeerrorresulttrue}
\end{figure}

Figure~\ref{fig:boreholeerrorresulttrue} shows the convergence of e and e$_{\text{T}}$ for the borehole problem when the hypothetical cost of the LF simulation is counted. Here, the MF-PCE method provides a good alternative for estimating the Sobol indices when one cannot afford the luxury of evaluating HF-PCE with higher sparse grid level. To reach a threshold value $10^{-3}$ and $10^{-5}$ for both e and e$_{\text{T}}$, MF-PCE only needs an HF sparse grid with one level lower than that of HF-PCE. The total computational cost is, of course, greater when the computational cost ratio is relatively high (i.e., RT=$\frac{1}{4}$); however, there is still a clear advantage of using MF-PCE over HF-PCE even when the RT is high for the borehole problem. In reaching a threshold value of $10^{-3}$, which we think is already a fine accuracy for engineering purpose, the MF-PCE method with RT=$\frac{1}{32},\frac{1}{16},\frac{1}{8}$ and $\frac{1}{4}$ need a computational cost equals to $17.6\%$, $20.8\%$, $27.07\%$, and $39.57\%$
of the cost needed by HF-PCE to reach the same threshold (i.e., HF-PCE with SG-3).

\begin{figure}[H]
\centering
\includegraphics[width=0.5\columnwidth]{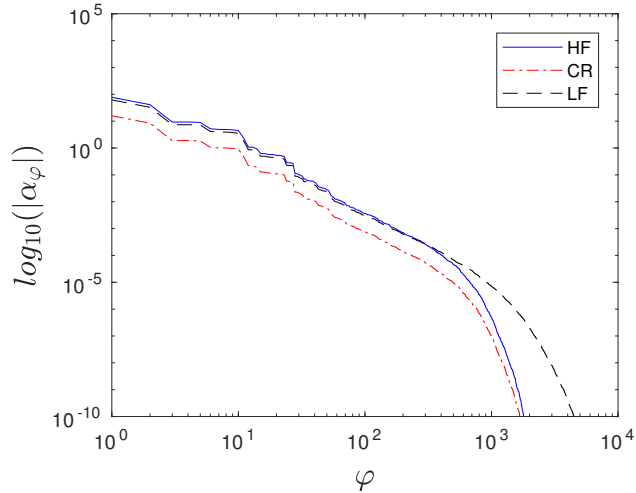}%
\caption{The decay of correction, HF-, and LF-PCE coefficients for the borehole function. Here, the correction function is less complex than HF- and LF-PCE, indicating a successful approximation.}
\label{fig:BOREHOLE_DECAY2_error}
\end{figure}

Figure~\ref{fig:BOREHOLE_DECAY2_error} shows the decay of PCE coefficients for the borehole problem. We use the SG-4-5 scheme to generate this plot. By analyzing Fig.~\ref{fig:BOREHOLE_DECAY2_error}, it is clear that the correction function is less complex than the HF- and LF-function, as indicated by its lower coefficients' magnitude and faster decay. Note that the correction PCE always seems to decay faster than LF-PCE due to the higher cardinality of LF-PCE. However, we can see that a significant portion of the correction PCE basis has lower coefficients values than that of LF-PCE. This less complexity indicates that it is easier to approximate the correction rather than the HF function, which leads to a more accurate approximation of the MF-PCE model compared to the standard HF-PCE model. Obviously, this leads to a more accurate approximation of the true Sobol indices. 

\subsection{Ishigami function}
\noindent The Ishigami function is a three-dimensional highly non-linear function which is frequently used as a benchmark problem for SA purpose~\cite{ishigami1990importance}. Here, we construct three LF representations of the Ishigami function to test the performance of MF-PCE in approximating the Sobol indices of a highly non-linear function. The standard form of the Ishigami function is defined as
\begin{equation}
f(\boldsymbol{\xi}) = \text{sin}(\xi_{1})+a\text{sin}^{2}(\xi_{2})+b\xi_{3}^{4}\text{sin}(\xi_{1}).
\end{equation}

The HF Ishigami function used in this paper, which is the standard expression of the Ishigami function, is expressed as
\begin{equation}
f_{h}(\boldsymbol{\xi}) = \text{sin}(\xi_{1})+7\text{sin}^{2}(\xi_{2})+0.1\xi_{3}^{4}\text{sin}(\xi_{1}).
\end{equation}
We then build three LF representations of the Ishigami function as follows: 
\begin{equation}
f_{l_{1}}(\boldsymbol{\xi}) = \text{sin}(\xi_{1})+7.3\text{sin}^{2}(\xi_{2})+0.08\xi_{3}^{4}\text{sin}(\xi_{1}),
\end{equation}
\begin{equation}
f_{l_{2}}(\boldsymbol{\xi}) = \text{sin}(\xi_{1})+7.3\text{sin}^{2}(\xi_{2})+0.04\xi_{3}^{4}\text{sin}(\xi_{1}),
\end{equation}
and
\begin{equation}
f_{l_{3}}(\boldsymbol{\xi}) = \text{sin}(\xi_{1})+7.3\text{sin}^{2}(\xi_{2})+0.04\xi_{3}^{4}\text{sin}(\xi_{1})+0.02.
\end{equation}

The r$^{2}_{lh}$ and MARE$_{lh}$ of the LF to the HF ishigami function are listed in Table~\ref{tbl:ishigamir2}. The distribution of the input random variables is Uniform$[-\pi,\pi]$ for all $\xi_{i}$. The three LF Ishigami models represent three scenarios of discrepancies between the HF and LF function. The 1\textsuperscript{st} LF function has the highest r$^{2}_{lh}$ correlation and lowest MARE$_{lh}$ while the 2\textsuperscript{nd} one has the opposite characteristics. On the other hand, the 3\textsuperscript{rd} LF model has exactly the same r$^{2}_{lh}$ as the 2\textsuperscript{nd} LF model but with higher MARE$_{lh}$. 

\begin{table}[h]
	\renewcommand{\arraystretch}{1.3}
	\caption{Statistical similarities between the LF and HF Ishigami functions.}
	\label{tbl:ishigamir2}
	\centering
	\begin{tabular}{ccc} \hline
		LF Model & r$^{2}_{lh}$ & MARE$_{lh}$ \\ \hline
		1 & 0.9875 & 0.4504 \\
		2 & 0.8826 & 1.0672\\
		3 & 0.8826 & 1.5587\\ \hline		
	\end{tabular}
\end{table}

The HF sparse grid level is varied from 1 to 6 with the fixed value of $q=2$. For the Ishigami function, the true Sobol indices can be obtained analytically, where the values are shown in Table~\ref{tbl:ishigamisobol}. These analytical Sobol Indices are obtained by decomposing the variance of the $f(\boldsymbol{\xi})$ as shown in Eq.~\ref{eq:ishigami1} as

\begin{equation}
\label{eq:ishigami1}
\begin{aligned}
D &= \frac{a^{2}}{8}+\frac{b\pi^{4}}{5}+\frac{b^{2}\pi^{8}}{18}+\frac{1}{2}, & &\\
D_{1}&=\frac{b\pi^{4}}{5}+\frac{b^{2}\pi^{8}}{50}+\frac{1}{2},~~D_{2}=\frac{a^{2}}{8}, ~~D_{3}=0,\\
D_{1,2}&=D_{2,3}=0,~~D_{1,3}=\frac{8b^{2}\pi^{8}}{225},~~D_{1,2,3}=0.
\end{aligned}
\end{equation}

\begin{figure}
\centering
\begin{subfigure}{.48\columnwidth}
\includegraphics[width=1\columnwidth]{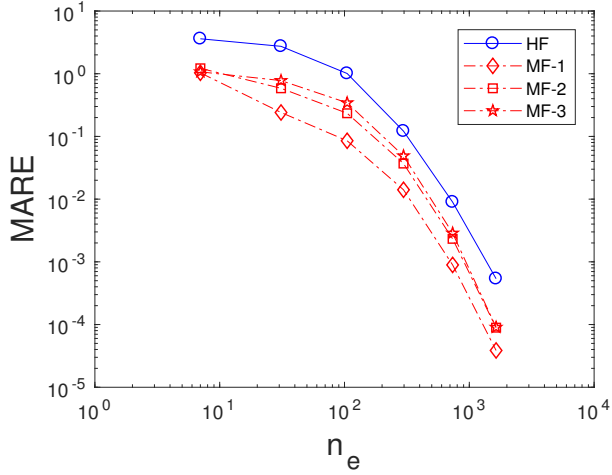}%
\caption{MARE.}%
\label{fig:ishigami_MARE_error}%
\end{subfigure}\hfill%
\begin{subfigure}{.48\columnwidth}
\includegraphics[width=1\columnwidth]{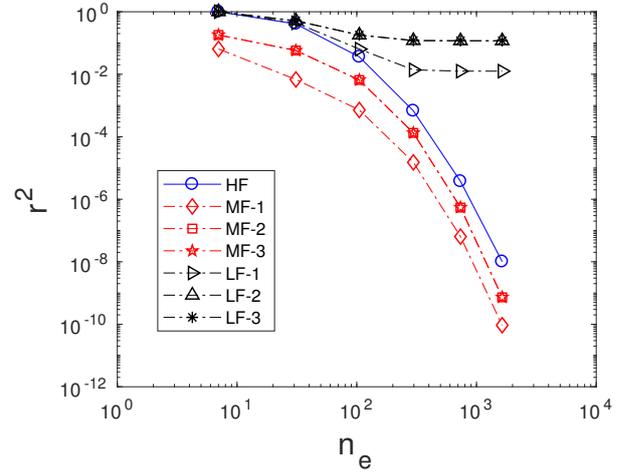}%
\caption{r$^{2}$}%
\label{fig:ishigami_R2_error}%
\end{subfigure}\hfill%
\begin{subfigure}{.48\columnwidth}
\includegraphics[width=1\columnwidth]{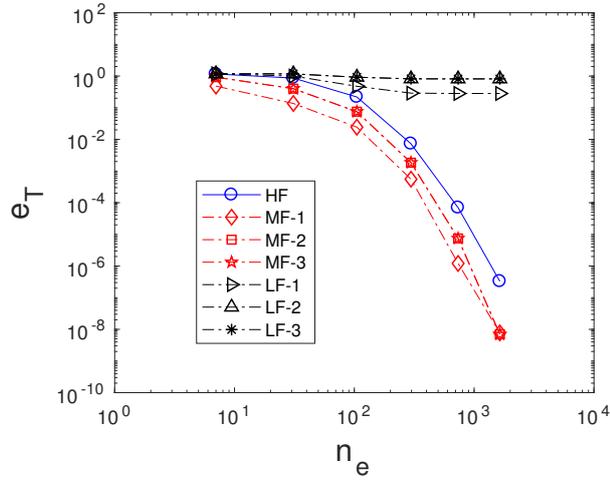}%
\caption{e$_{\text{T}}$.}%
\label{fig:ishigami_TS_error}%
\end{subfigure}\hfill%
\begin{subfigure}{.48\columnwidth}
\includegraphics[width=1\columnwidth]{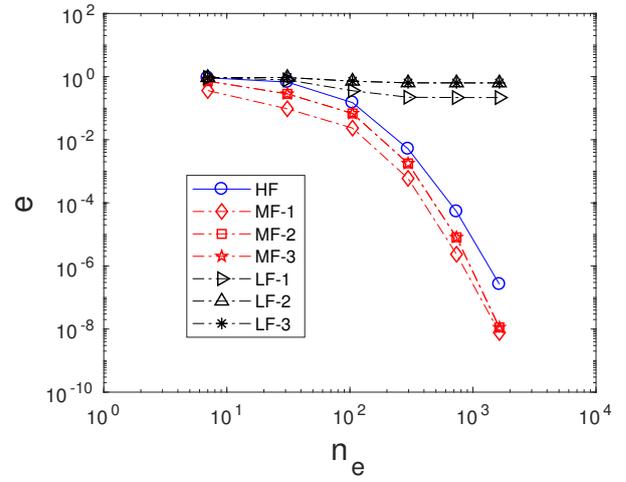}%
\caption{e.}%
\label{fig:ishigami_AS_error}%
\end{subfigure}\hfill%
\caption{Convergence of error metrics for the Ishigami function with respect to the number of function evaluations. The MF-PCE scheme with the first LF model estimates the Sobol indices better than the other methods. Here, solely LF-PCE cannot be relied on to estimate the Sobol indices since the correlation is not near one.}
\label{fig:ishigamiresult}
\end{figure}

The results, as shown in Fig.~\ref{fig:ishigamiresult}, indicate that all MF-PCE schemes are able to reduce the Sobol indices error compared to HF-PCE. It can also be seen that the convergence trend for all error metrics is roughly similar. The MF-PCE method with the 1\textsuperscript{st} LF model produces the MF model with the lowest error of all, which is reasonable due to its high r$^{2}_{lh}$ with the HF function. Furthermore, there is no significant difference between the convergence error of the MF scheme with the 2\textsuperscript{nd} and 3\textsuperscript{rd} LF models. Although the MARE$_{lh}$ of the 3\textsuperscript{rd} LF model is higher than the 2\textsuperscript{nd} model (See Fig.~\ref{fig:ishigami_MARE_error}), the trend of both functions is similar to each other which results in almost the same quality of the MF-PCE model. All LF-PCEs without correction (see Fig.~\ref{fig:ishigami_R2_error}), as it is already obvious, cannot accurately estimate the r$^{2}$ of the true Ishigami function. Thus, LF-PCE fails to estimate the true Sobol Indices (See Fig.~\ref{fig:ishigami_TS_error} and~\ref{fig:ishigami_AS_error}). This is mainly due to the r$^{2}_{lh}$ correlation which is not very close to one, unlike the borehole case as explained before. In cases like this, the LF model without correction should not be trusted as the representation of the HF function to estimate the Sobol indices. However, when the LF model is used within the MF-PCE framework, it can be efficiently used to estimate the Sobol indices of the true HF function with lower computational cost. This is very useful if the computational cost ratio of the HF to the LF function is very high.

\begin{figure}[H]
\centering
\begin{subfigure}{.48\columnwidth}
\includegraphics[width=1\columnwidth]{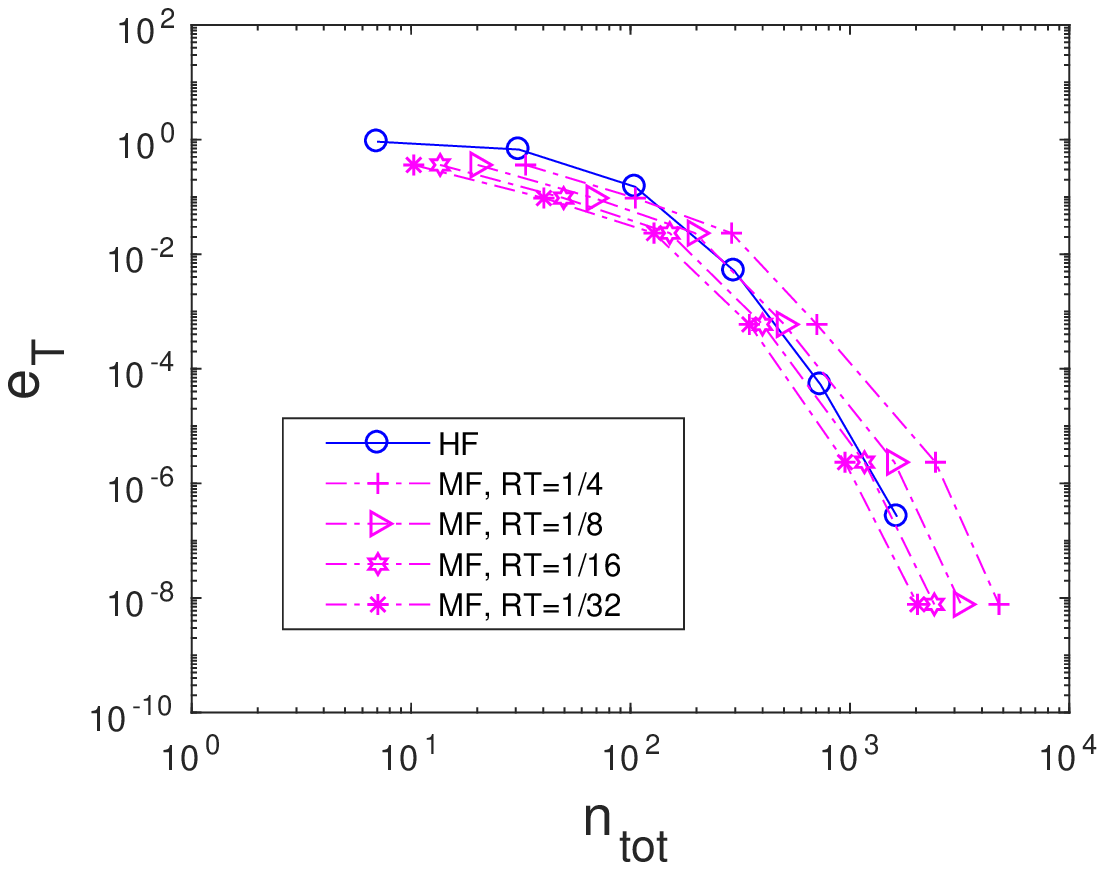}%
\caption{e$_{\text{T}}$.}%
\label{fig:ISHIGAMI_TS_error_ntot}%
\end{subfigure}\hfill%
\begin{subfigure}{.48\columnwidth}
\includegraphics[width=1\columnwidth]{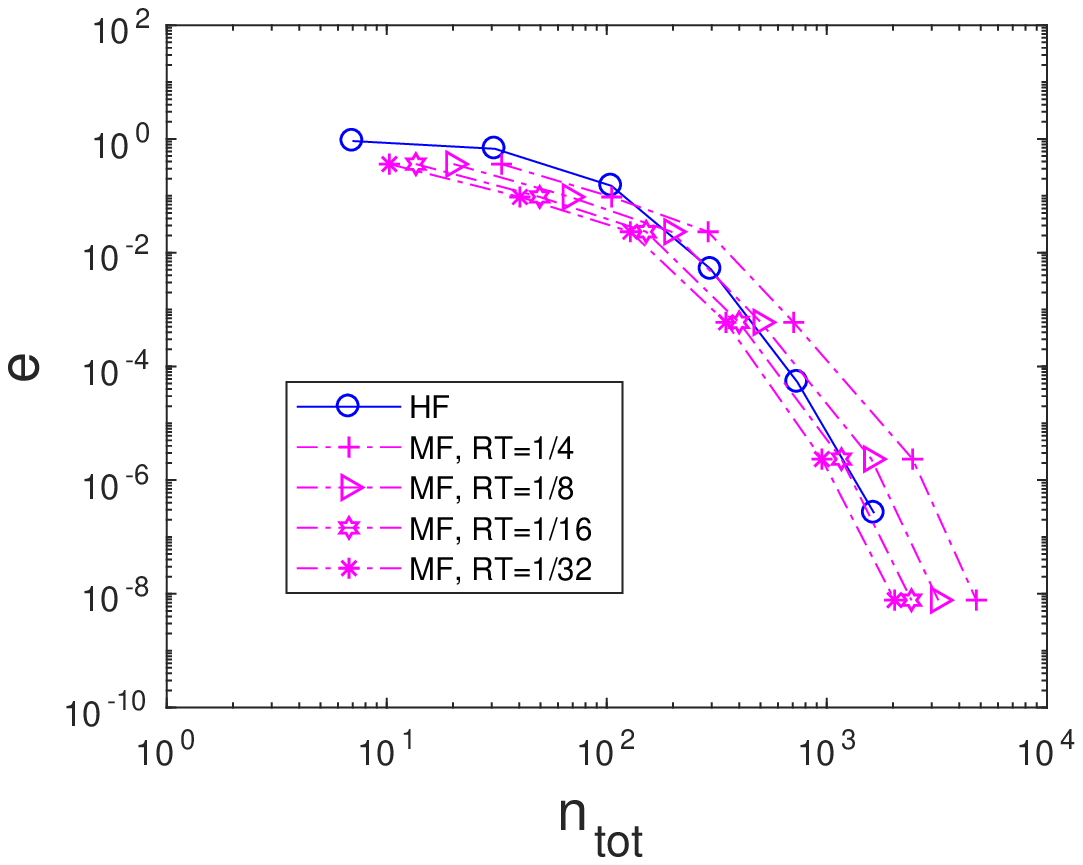}%
\caption{e.}%
\label{fig:ISHIGAMI_AS_error_ntot}%
\end{subfigure}\hfill%
\caption{Convergence of error metrics for the Ishigami function with respect to the total simulation cost assuming a hypothetical LF model cost. Only MF-PCE with the first LF model is shown due to its highest accuracy over the other LF models.}
\label{fig:ishigamierrorresulttrue}
\end{figure}

The convergence of error metrics with the hypothetical actual cost is shown in Fig.~\ref{fig:ishigamierrorresulttrue}. Here, only MF-PCE with the first LF model is shown since it is the best LF model available for the Ishigami function. By observing the plot, we can see that MF-PCE fails to yield a significant beneficial effect when the computational cost ratio is equal to $1/4$. We can see that the computational cost of MF-PCE with RT$=1/4$ is $96.16\%$ of HF-PCE to reach a threshold value of $10^{-3}$ for both e and e$_{\text{T}}$, which is actually not a significant cost reduction. On the other hand, if the computational cost ratio is assumed to be $1/32,1/16$ and $1/8$, the cost to reach this threshold are $47.38\%$, $54.35\%$, and $68.28\%$ of the cost needed by HF-PCE, respectively. When the threshold is set to be very strict, that is, $10^{-5}$, the cost of MF-PCE to reach this threshold with RT$=1/32$, $1/16,1/8$ and $1/4$ are $57.95\%$, $71.06\%$, $97.29\%$ and $149.74\%$ of HF-PCE, respectively. With this very strict threshold, the cost reduction becomes significant only when the computational cost ratio equals to $1/16$ or $1/32$. Conversely, when the computational cost ratio equals to $1/4$, MF-PCE requires more functions evaluations than that of HF-PCE to reach this very high accuracy.

\begin{figure}
\centering
\begin{subfigure}{.48\columnwidth}
\includegraphics[width=1\columnwidth]{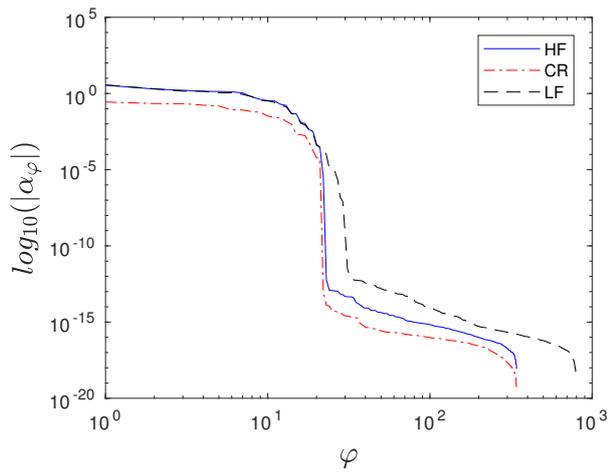}%
\caption{1\textsuperscript{st} LF model.}%
\label{fig:ISHIGAMI_DECAY_MOD1_error}%
\end{subfigure}\hfill%
\begin{subfigure}{.48\columnwidth}
\includegraphics[width=1\columnwidth]{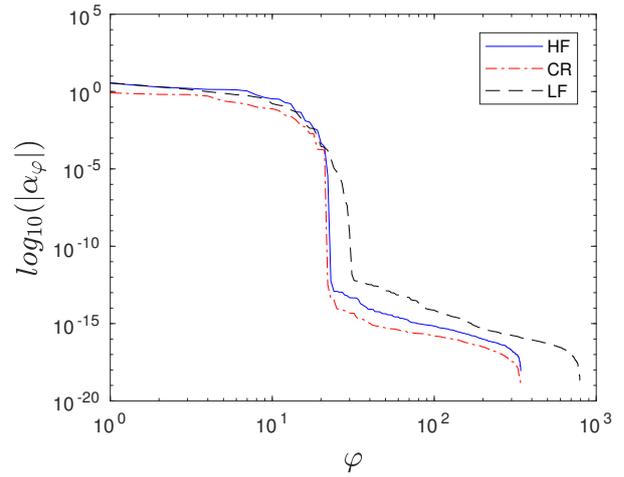}%
\caption{2\textsuperscript{nd} LF model.}%
\label{fig:ISHIGAMI_DECAY_MOD2_error}%
\end{subfigure}\hfill%
\begin{subfigure}{.48\columnwidth}
\includegraphics[width=1\columnwidth]{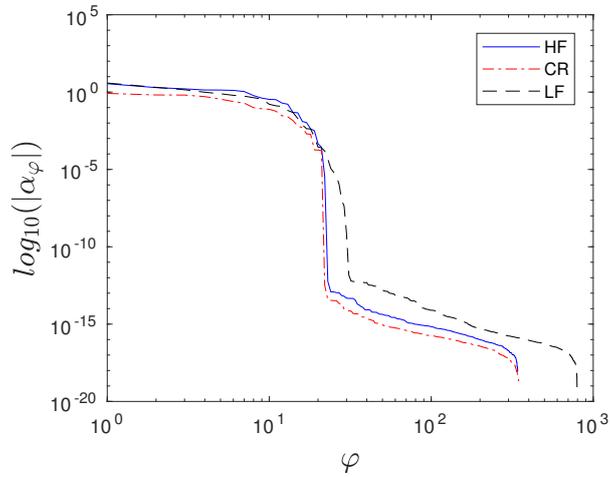}%
\caption{3\textsuperscript{rd} LF model.}%
\label{fig:ISHIGAMI_DECAY_MOD3_error}%
\end{subfigure}\hfill%
\caption{The decay of correction, HF-, and LF-PCE coefficients for the Ishigami function. The correction functions from all the three LF models are less complex (i.e., lower magnitude and faster decay) than their corresponding HF- and LF-PCE, with the gap is wider for the first LF model.}
\label{fig:ISHIGAMI_DECAY}
\end{figure}

The decay of PCE coefficients, as shown in Fig.~\ref{fig:ISHIGAMI_DECAY}, clearly indicates that the coefficients of all correction PCEs decay faster than those of LF- and HF-PCE for all the LF models. This means that the correction functions are all less complex than the LF one. By visual investigation, it can be clearly seen that the gap between the HF and correction PCE model is wider for MF-PCE with the 1\textsuperscript{st} model than that of the others. The wider this gap, the less complex the correction to the LF function, which results in a more successful MF model in estimating the Sobol indices.

\begin{table}[h]
\renewcommand{\arraystretch}{1.3}
\caption{First order and total Sobol indices of the HF Ishigami function.}
\label{tbl:ishigamisobol}
\centering
\begin{tabular}{cc} \hline
Index & Analytical  \\ \hline
$SU_{1}$ & 0.3138  \\
$SU_{2}$ & 0.4424  \\
$SU_{3}$ & 0  \\
$SU_{12}$ & 0  \\
$SU_{13}$ & 0.2436  \\
$SU_{23}$ & 0  \\
$SU_{123}$ & 0  \\
$SU_{1}^{T}$ & 0.5574 \\
$SU_{2}^{T}$ & 0.4424  \\
$SU_{3}^{T}$ & 0.2436  \\ \hline

\end{tabular}
\end{table}

\subsection{Short column problem}
The third algebraic test problem is a short column problem with five LF models~\cite{ng2012multifidelity}. Furthermore, this problem also involves random variables with a non-uniform distribution.

The HF response for the short column problem is defined as
\begin{equation}
f_{h}(\boldsymbol{\xi}) = 1- \frac{4M}{bh^{2}Y}-(\frac{P}{bhY})^{2}.
\end{equation}

The LF responses are defined as~\cite{ng2012multifidelity}
\begin{equation}
f_{l_{1}}(\boldsymbol{\xi}) = 1- \frac{4P}{bh^{2}Y}-(\frac{P}{bhY})^{2},
\end{equation}
\begin{equation}
f_{l_{2}}(\boldsymbol{\xi}) = 1- \frac{4M}{bh^{2}Y}-(\frac{M}{bhY})^{2},
\end{equation}
\begin{equation}
f_{l_{3}}(\boldsymbol{\xi}) = 1- \frac{4M}{bh^{2}Y}-(\frac{P}{bhY})^{2}-\frac{4(P-M)}{bhY},
\end{equation}
\begin{equation}
f_{l_{4}}(\boldsymbol{\xi}) = 1- \frac{4M}{bh^{2}Y}-(\frac{P}{bhY})^{2}-\frac{0.4(P-M)}{bhY},
\end{equation}
and
\begin{equation}
f_{l_{5}}(\boldsymbol{\xi}) = 1- \frac{4M}{bh^{2}Y}-(\frac{P}{bhY})^{2}-\frac{40(P-M)}{bhY}.
\end{equation}
where $\boldsymbol{\xi} = (b,h,P,M,Y)$ with the distribution listed in Table~\ref{tbl:shortcol}. The five LF short column functions have various values of r$^{2}_{lh}$ correlation and MARE$_{lh}$ to the HF one, as shown in Table~\ref{tbl:shortcolumncor}. As it can be seen from this table, there are some LF models that have high r$^{2}_{lh}$ and low MARE$_{lh}$ (model 1, for example) and LF models with low correlation value and very large MARE$_{lh}$ (model 3, for example). We aim to further investigate the criteria of which LF model can yield better accuracy in estimating Sobol indices when employed in the MF-PCE framework. The sparse grid level offset $q$ is set to 2 for this problem.

\begin{table}[h]
	\centering
	\begin{tabular}{cc} \hline
		Random variable & Probability distribution \\ \hline 
		$b$  & Uniform [5,15]   \\ 
		$h$  &  Uniform [15,25]  \\ 
		$P$  & Normal [500,100]  \\ 
		$M$  & Normal [2000,400]  \\ 
		$Y$   & Normal [5,0.5]   \\ \hline
	\end{tabular}
	\caption{Random variable definition and distributions for the short column test case.}
	\label{tbl:shortcol}
\end{table}

\begin{table}[h]
\centering
\begin{tabular}{ccc} \hline
LF model & r$_{lh}^{2}$ & MARE$_{lh}$ \\ \hline
1 & 0.9234   & 8.6926 \\
2 & 0.7612 & 126.6604\\
3 & 0.4780 & 154.7131\\
4 & 0.6931 & 15.4713\\
5 & 0.5914 & 1547.13\\
\hline
\end{tabular}
\caption{Correlation and MARE$_{lh}$ values between the LF and HF short column responses.}
\label{tbl:shortcolumncor}
\end{table}

\begin{table}[h]
\renewcommand{\arraystretch}{1.3}
\caption{First order and total Sobol indices of the HF short column function.}
\label{tbl:shocolsobol}
\centering
\begin{tabular}{cc} \hline
Index & MCS  \\ \hline
$SU_{1}$ & 0.5240  \\
$SU_{2}$ & 0.2059  \\
$SU_{3}$ & 0.0696 \\
$SU_{4}$ & 0.0284  \\
$SU_{5}$ & 0.0532  \\
$SU_{1}^{T}$ & 0.6257 \\
$SU_{2}^{T}$ & 0.2708  \\
$SU_{3}^{T}$ & 0.1143  \\
$SU_{4}^{T}$ & 0.0346  \\
$SU_{5}^{T}$ & 0.0799 \\ \hline

\end{tabular}
\end{table}

\begin{figure}[H]
\centering
\begin{subfigure}{.48\columnwidth}
\includegraphics[width=1\columnwidth]{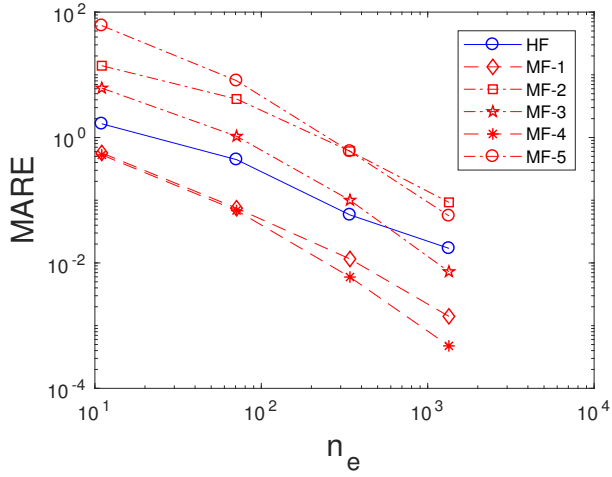}%
\caption{MARE.}%
\label{fig:SHORTCOL_MARE_error}%
\end{subfigure}\hfill%
\begin{subfigure}{.48\columnwidth}
\includegraphics[width=1\columnwidth]{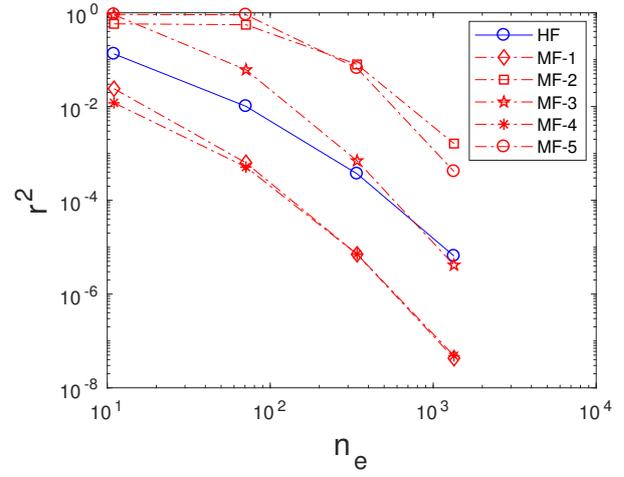}%
\caption{r\textsuperscript{2}.}%
\label{fig:SHORTCOL_R2_error}%
\end{subfigure}\hfill%
\begin{subfigure}{.48\columnwidth}
\includegraphics[width=1\columnwidth]{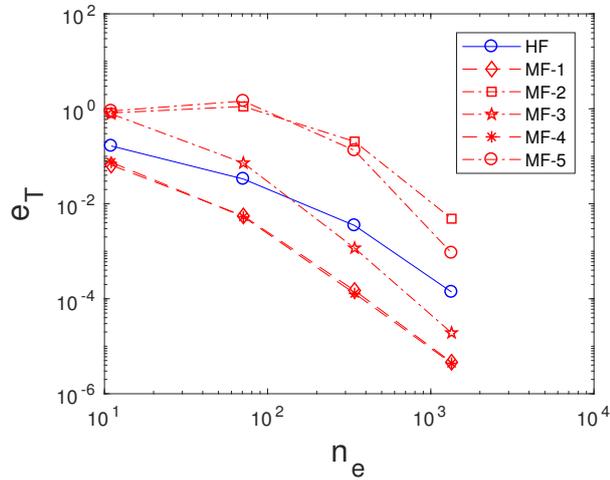}%
\caption{e$^\text{T}$.}%
\label{fig:SHORTCOL_TS_error}%
\end{subfigure}\hfill%
\begin{subfigure}{.48\columnwidth}
\includegraphics[width=1\columnwidth]{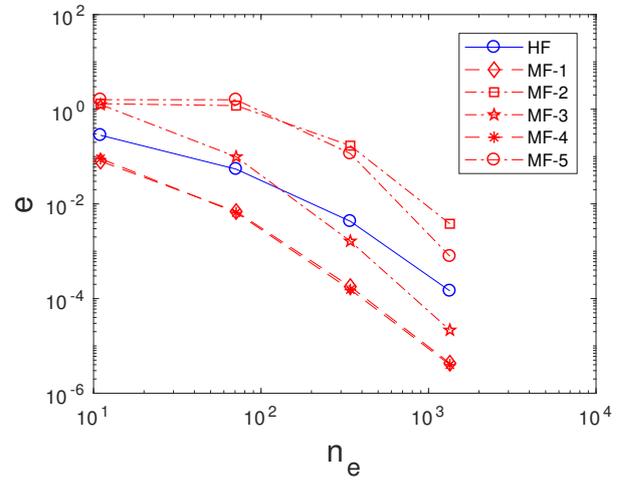}%
\caption{e.}%
\label{fig:SHORTCOL_AS_error}%
\end{subfigure}\hfill%
\caption{Convergence of error metrics for the short column function with respect to the number of function evaluations. The MF-PCE scheme with the first and the fourth LF models consistently outperform HF-PCE for all sparse grid levels.}
\label{fig:shortcolerror}
\end{figure}

The convergence of the error metrics for the short column problem is shown in Fig.~\ref{fig:shortcolerror}. By closer investigation of this figure, we observe that only the 1\textsuperscript{st} and 4\textsuperscript{th} LF functions are able to successfully reduce the error of Sobol indices relative to HF-PCE for all levels of sparse grid. As expected, the decrement of MARE and r$^{2}$ corresponds to a better estimation of Sobol indices. However, it is not always the case as it can be observed from the result of MF-PCE using the 3\textsuperscript{rd} LF model which produces a better estimate of Sobol indices than that of HF-PCE from HF sparse grid level of 3. The 2\textsuperscript{nd} and 5\textsuperscript{th} LF models produce MF models with lower quality than those of HF-PCE in terms of estimating the Sobol indices. 

\begin{figure}[H]
\centering
\begin{subfigure}{.48\columnwidth}
\includegraphics[width=1\columnwidth]{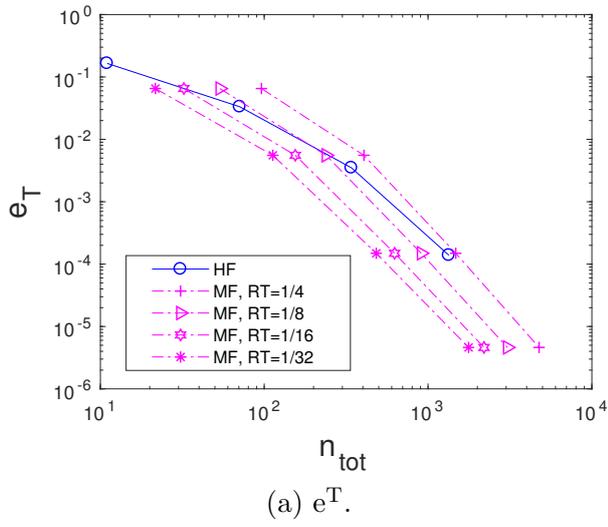}%
\caption{e$^\text{T}$.}%
\label{fig:SHOCOL_TS_error_ntot}%
\end{subfigure}\hfill%
\begin{subfigure}{.48\columnwidth}
\includegraphics[width=1\columnwidth]{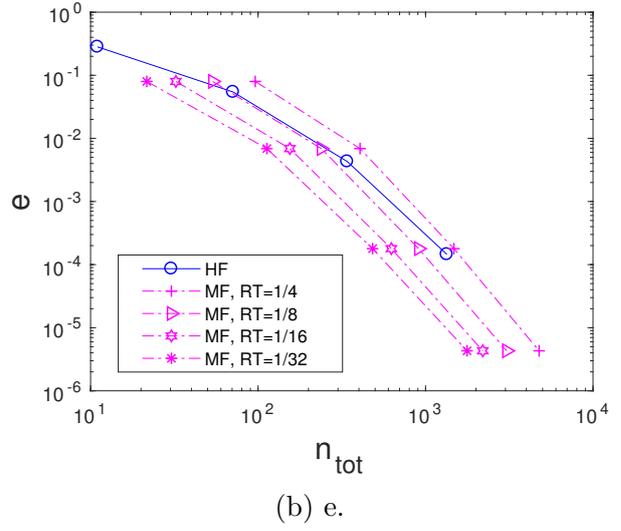}%
\caption{e.}%
\label{fig:SHOCOL_AS_error_ntot}%
\end{subfigure}\hfill%
\caption{Convergence of error metrics for the short column function with respect to the total simulation cost assuming a hypothetical LF model cost. Only the first LF model is used since it is the best available LF model for MF approximation.}
\label{fig:shocolerrorresulttrue}
\end{figure}

Figure~\ref{fig:shocolerrorresulttrue} depicts the convergence of error metrics for the short column function with the hypothetical cost of the LF simulation is taken into account. It is clear from this figure that when the computational cost ratio is equal to 1/4, there is no advantage gained by utilizing MF-PCE. This is because  MF-PCE with RT=1/4 needs 110\% of the computational cost needed by HF-PCE to reach the e and e$_{\text{T}}$ below $10^{-3}$. On the other hand, we observe that MF-PCE with RT=1/8 requires 67.8\% of the HF-PCE cost to reach a similar accuracy, which is quite a noteworthy cost reduction for reaching an engineering accuracy. The case with RT=1/16 and RT=1/32 require 46.61\% and 36.02\% of the HF-PCE cost to reach the engineering accuracy of $10^{-3}$. 

By only investigating statistical similarities between the HF and LF models, it is hard to determine whether an MF-PCE with a certain LF-model could successfully estimate the true Sobol indices or not. It is clear for the 1\textsuperscript{st} model, but not for the 4\textsuperscript{th} model since its r$^{2}_{lh}$ correlation is low and the MARE$_{lh}$ is relatively high. It becomes much more difficult to make this inference for the 2\textsuperscript{nd} LF model (its correlation is moderate while the MARE$_{lh}$ is very high). Although for some cases it is obvious to see the adequacy of the LF model for assisting MF-PCE, inspecting the PCE coefficients decay is the proper way to do this kind of analysis.

\begin{figure}[H]
\centering
\begin{subfigure}{.46\columnwidth}
\includegraphics[width=1\columnwidth]{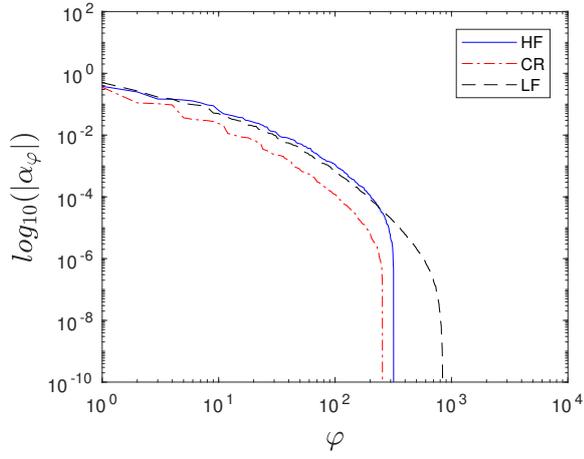}%
\caption{1\textsuperscript{st} model.}%
\label{fig:SHORTCOL_DECAY1_SIerror}%
\end{subfigure}\hfill%
\begin{subfigure}{.46\columnwidth}
\includegraphics[width=1\columnwidth]{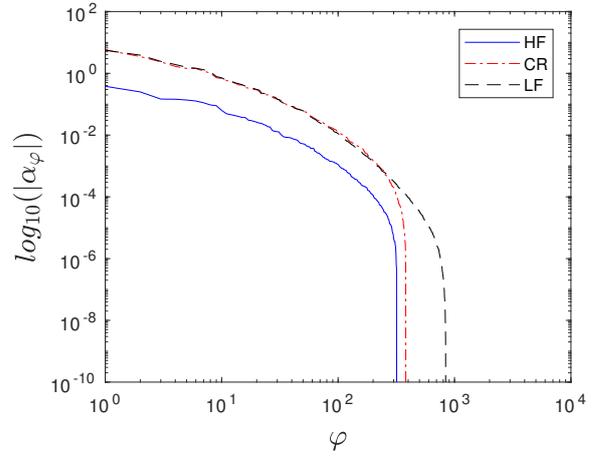}%
\caption{2\textsuperscript{nd} model.}%
\label{fig:SHORTCOL_DECAY2_SIerror}%
\end{subfigure}\hfill%
\begin{subfigure}{.46\columnwidth}
\includegraphics[width=1\columnwidth]{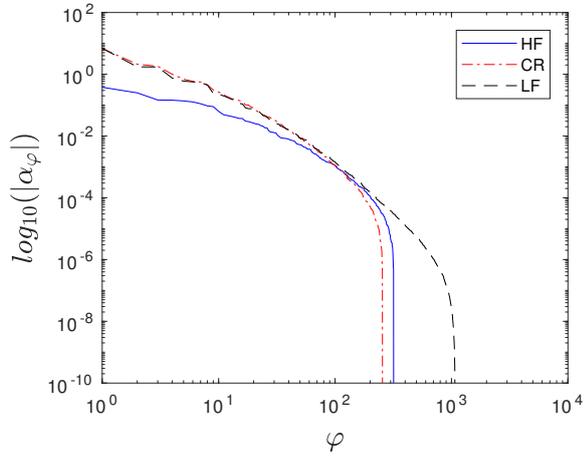}%
\caption{3\textsuperscript{rd} model.}%
\label{fig:SHORTCOL_DECAY3_SIerror}%
\end{subfigure}\hfill%
\begin{subfigure}{.46\columnwidth}
\includegraphics[width=1\columnwidth]{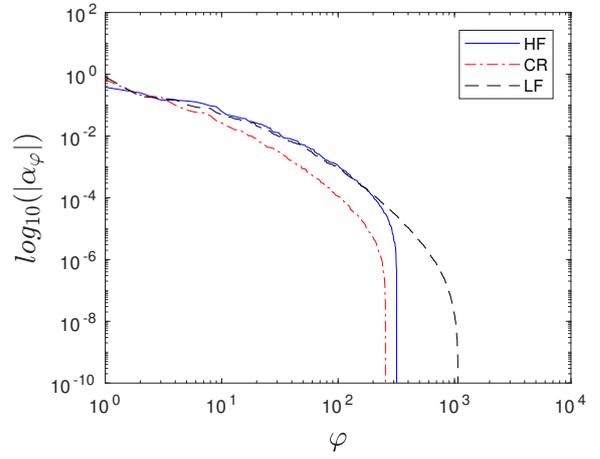}%
\caption{4\textsuperscript{th} model.}%
\label{fig:SHORTCOL_DECAY4_SIerror}%
\end{subfigure}\hfill%
\begin{subfigure}{.46\columnwidth}
\includegraphics[width=1\columnwidth]{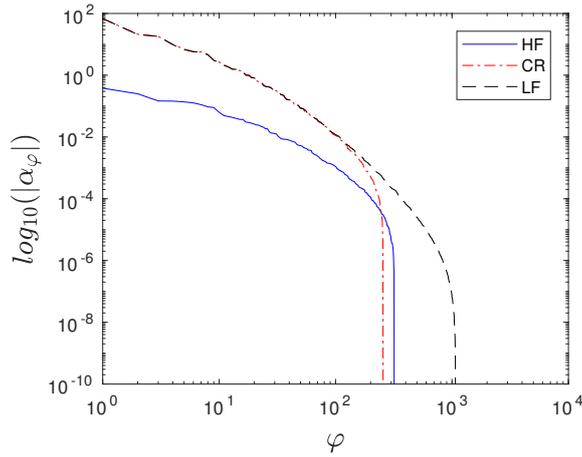}%
\caption{5\textsuperscript{th} model.}%
\label{fig:SHORTCOL_DECAY5_SIerror}%
\end{subfigure}\hfill%
\caption{The decay of the correction, HF-, and LF-PCE coefficients for the short column function with various LF models. The correction PCE's coefficients of the first and the fourth LF models decay faster than those of the corresponding HF-PCE, indicating a successful MF model.}
\label{fig:shortcoldecay}
\end{figure}

The decay of the PCE coefficients for all LF models are shown in Fig.~\ref{fig:shortcoldecay}. All PCE coefficients are obtained using level 4 correction sparse grids (means level 6 LF sparse grid), which are adequate enough to obtain good convergence of the Sobol indices. It is very clear from this figure that the behavior of coefficients decay for the HF, LF, and correction PCE are similar for unsuccessful MF models. Here, we can see that the magnitude of the correction PCEs' coefficients is much larger and it also decay slower than that of HF-PCE. Moreover, the coefficients of the LF and correction PCE decay similarly to each other for the 2\textsuperscript{nd}, 3\textsuperscript{rd}, and 5\textsuperscript{th} LF models. Such similar decay means that the trend of correction PCE almost mimics that of LF-PCE, indicating that the correction PCE is not less complex than LF- and HF-PCE. The MF-PCE scheme with the 3\textsuperscript{rd} LF model is a special case, where the convergence of e and e$_{\text{T}}$ shows that it slightly outperformed HF-PCE when correction sparse grid with level 3 and 4 are used. However, it is not suggested to use the LF model like the 3\textsuperscript{rd} model in a real application. This is because we cannot trust the adequacy of the 3\textsuperscript{rd} model in aiding MF-PCE when investigating the PCE coefficients decay. Indeed, with a low level sparse grid, MF-PCE with the 3\textsuperscript{rd} LF model fails to estimate the Sobol indices accurately. On the other hand, the decay trend of MF-PCE with the 1\textsuperscript{st} and 4\textsuperscript{th} models suggests that the MF approximation is better than the HF one. The gap between the decay of the correction and HF-, LF-PCE is visible for these two LF models, indicating the less complexity of the correction-PCE model. This difference especially can be clearly observed when MF-PCE with the 1\textsuperscript{st} LF model is compared with the 4\textsuperscript{th} LF model. In addition, if we pay a close attention to the decay of the PCE coefficients for the 4\textsuperscript{th} model, the magnitude of the correction PCE's first coefficient is larger than that of HF-PCE (i.e., the mean is higher). However, its correction PCE's coefficients decay faster than that of HF-PCE. In the light of these results, we observe that it is the rate of decay of the coefficients that is important for a successful MF-PCE model and not the difference between the mean value of the correction and HF-PCE. Moreover, it is also better to look at the difference between correction PCE and both HF- and LF- PCE.

\subsection{Transonic airfoil in inviscid flow}
The final test is the SA of a transonic airfoil in inviscid flow. The open source CFD code of SU$^{2}$ was used to solve the Euler equation in order to obtain the aerodynamic coefficients~\cite{palacios2013stanford}. The airfoil model in this case is assumed to be subjected to geometrical uncertainties. Here, we used the RAE 2822 airfoil which was parameterized via PARSEC method to find the variables that describe the datum shape~\cite{sobieczky1999parametric}. Note that since the datum shape is an approximation of the original one, there is an error between the coordinate of the true and the approximated shape. However, our goal here is to demonstrate the capability of MF-PCE in performing SA, and not to measure the randomness of the original airfoil itself; hence, this test case is appropriate enough for the purpose of demonstrating an SA algorithm. For this problem, the random output of interest is the lift-to-drag ratio ($L/D$), under the fix flight condition of Mach number and angle of attack of 0.73 and 2$^{0}$, respectively. 
 
\begin{figure}[H]
 \centering
 \includegraphics[width=0.5\columnwidth]{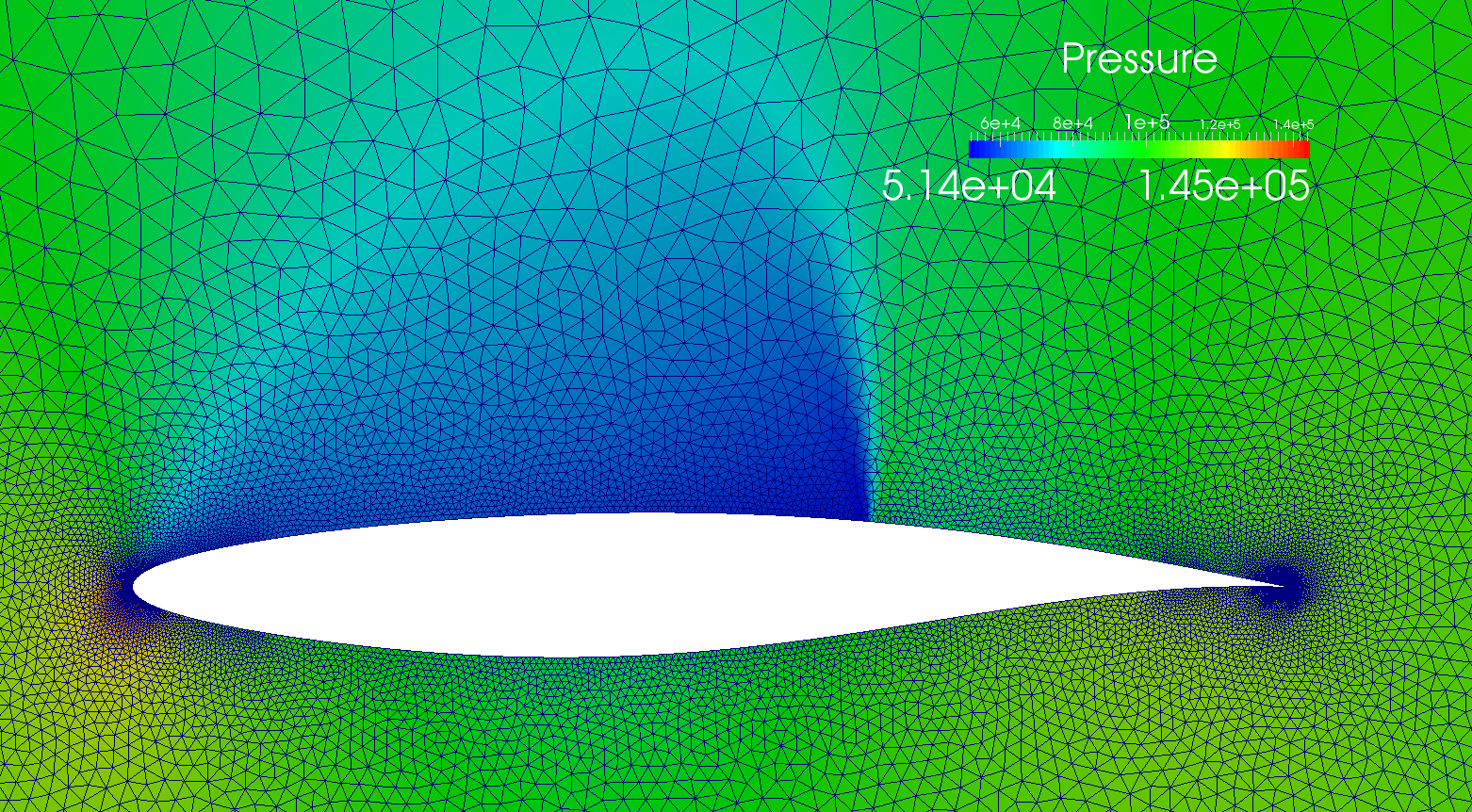}%
 \caption{An example of one CFD realization for the inviscid transonic airfoil case.}
 \label{fig:mesh_fine_SA_1.png}
 \end{figure}

\begin{figure}[H]
  \centering
  \includegraphics[width=0.5\columnwidth]{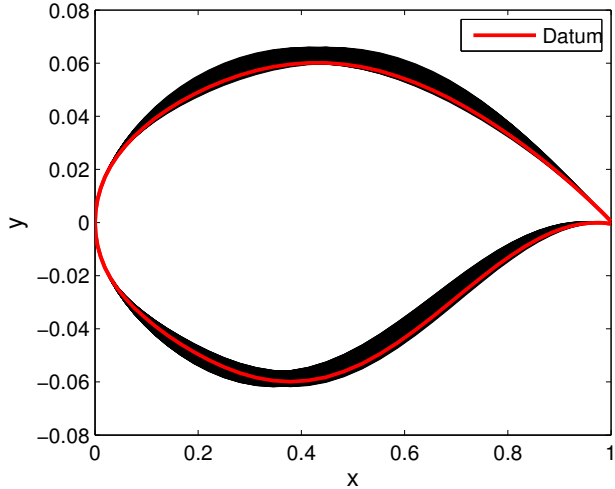}%
  \caption{The datum airfoil shape (red line) and the ensemble of various realizations of the uncertain shape (black lines) for the inviscid transonic airfoil case.}
  \label{fig:PARSEC_RANDOM}
  \end{figure}

After the suitable PARSEC parameters were found, we assume that there were five uncertain parameters, that is, the $x$ and $y$ coordinate of the lower crest (i.e., $x_{lo}$ and $y_{lo}$, respectively),  $x$ and $y$ coordinate of the upper crest (i.e., $x_{up}$ and $y_{up}$, respectively), and trailing edge angle (i.e., $\alpha_{te}$), with the distribution listed in Table~\ref{tbl:parseccase}. An example of one realization evaluated by the CFD solver is shown in Fig.~\ref{fig:mesh_fine_SA_1.png}. The datum shape and several realizations of the uncertain shape are shown in Fig.~\ref{fig:PARSEC_RANDOM}. For this problem, we used the fully and partially converged simulations as the HF and LF simulations, respectively. The criterion for the fully converged simulation is the residual of the drag that reaches $10^{-6}$, which on average requires 500 iterations to reach this criterion. On the other hand, the iteration for the partially converged simulation to be used as the LF simulations is set to 100, which means that the computational cost ratio between the LF and HF simulation is 0.2 for this case. This fully and partially converged simulation strategy has been employed for optimization~\cite{forrester2006optimization, branke2017efficient}, surrogate model construction~\cite{courrier2014use}, and UQ~\cite{palar2016multi}. One advantage of this strategy is that some LF samples are evaluated together with the HF samples, which leads to further reduction in the computational cost.
 
 Since the statistical moments are typically also the quantities of interest in real-world applications, we plotted the convergence of statistical moments together with the Sobol indices. For reference value, an HF-PCE with level 4 sparse grid is used to obtain the reference total Sobol indices and statistical moments, as shown in Table~\ref{tbl:parsecsobol}. Shown in this table are also the Sobol indices obtained by LF-PCE with level 4 sparse grid. By ranking the total sensitivity indices, it is found that $y_{up}$ is the most important variable followed by $x_{up}$ and $y_{lo}$, while the contribution of $x_{lo}$ and $\alpha_{te}$ is very small but not negligible. The sum of the total Sobol indices is 1.015, which means that the largest contributors to the total variance are the first order sensitivity indices with small interaction between variables. Although the Sobol indices obtained by LF-PCE seem close to the true Sobol indices, they are not accurate representations of the true indices. Moreover, the standard deviation of the HF- and LF-PCE differs by 1.78\%. This means that correction of LF-PCE must be applied within MF-PCE framework.

\begin{table}[h]
	\centering
	\begin{tabular}{ccc} \hline
		No. & Random variable & Probability distribution \\ \hline 
		1 & $x_{up}$  & Uniform [0.4415, 0.4549]   \\ 
		2 & $y_{up}$  &  Uniform [0.060, 0.066]  \\ 
		3 & $x_{lo}$  & Uniform [0.344, 0.3800]  \\ 
		4& $y_{lo}$  & Uniform [-0.0618, -0.0560]  \\ 
		5&$\alpha_{te}$   & Uniform [-0.1182,-0.1070]   \\ \hline
	\end{tabular}
	\caption{Random variable definition and distributions for the inviscid transonic airfoil test case.}
	\label{tbl:parseccase}
\end{table}

\begin{table}[h]
	\centering
	\begin{tabular}{ccc} \hline
		Sensitivity index & HF, SG-4  & LF, SG-4   \\ \hline
		$S^{T}_{1}$	  & 0.3207   & 0.3136   \\ 
		$S^{T}_{2}$	  & 0.6964    & 0.6970 	  \\ 
		$S^{T}_{3}$	  & 0.0044    & 0.0037     \\ 
		$S^{T}_{4}$	  & 0.0214    & 0.0198    \\ 
		$S^{T}_{5}$	  & 0.0085    & 0.0084 	  \\ 
		\hline
		Mean		  & 121.9250    & 121.5645 \\ 
		St. D.		  & 39.6901    & 37.3553 \\ \hline
	\end{tabular}
	\caption{Sobol indices obtained by HF- and LF-PCE with high level sparse grid for the inviscid transonic airfoil problem.}
	\label{tbl:parsecsobol}
\end{table}

\begin{figure}[H]
\centering
\begin{subfigure}{.48\columnwidth}
\includegraphics[width=1\columnwidth]{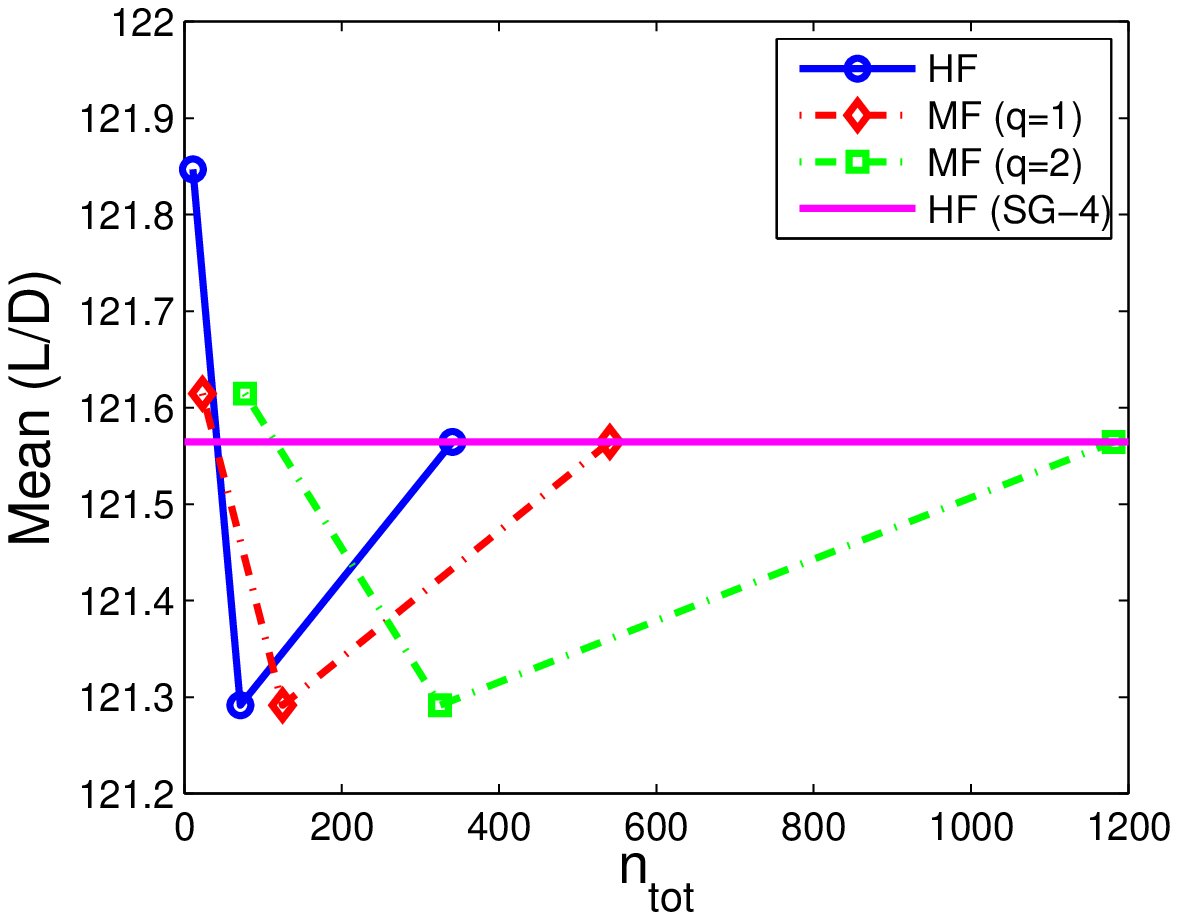}%
\caption{Mean.}%
\label{fig:PARSEC_SP_mean}%
\end{subfigure}\hfill%
\begin{subfigure}{.48\columnwidth}
\includegraphics[width=1\columnwidth]{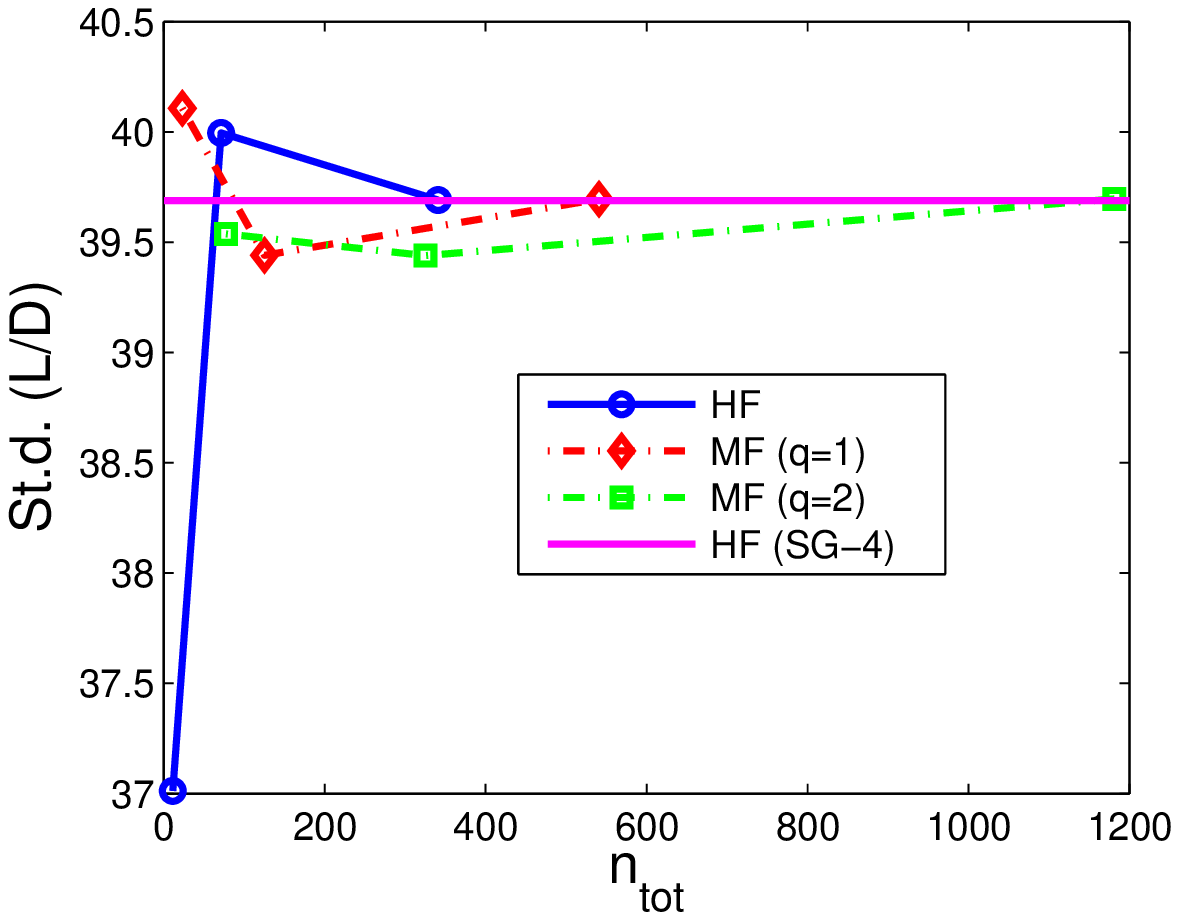}%
\caption{Standard deviation.}%
\label{fig:PARSEC_SP_std}%
\end{subfigure}\hfill%
\begin{subfigure}{.48\columnwidth}
\includegraphics[width=1\columnwidth]{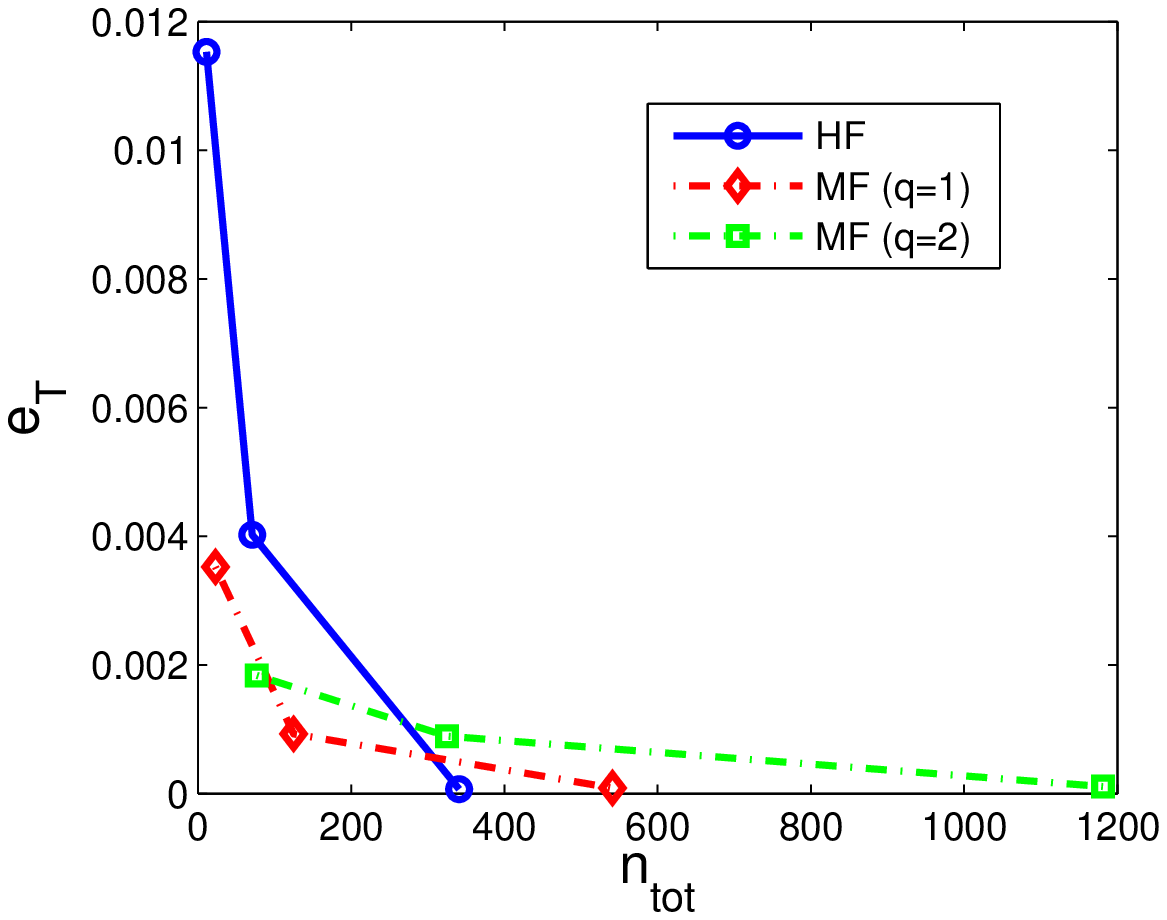}%
\caption{e$_{\text{T}}$}%
\label{fig:PARSEC_SP_ET}%
\end{subfigure}\hfill%
\caption{Convergence of statistical moments and total Sobol indices for the inviscid transonic airfoil problem. The advantage of utilizing MF-PCE is more evident when it is used to estimate the total Sobol indices as shown by the faster convergence of MF-PCE in approaching engineering accuracy.}
\label{fig:PARSECmoments}
\end{figure}

\begin{figure}[H]
\centering
\begin{subfigure}{.45\columnwidth}
\includegraphics[width=1\columnwidth]{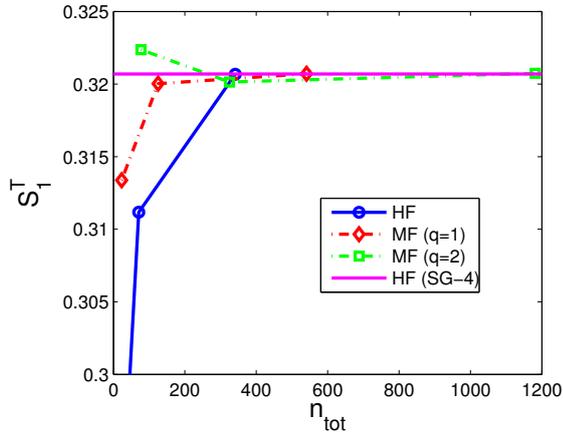}%
\caption{$x_{up}$.}%
\label{fig:PARSEC_SP_ST1}%
\end{subfigure}\hfill%
\begin{subfigure}{.45\columnwidth}
\includegraphics[width=1\columnwidth]{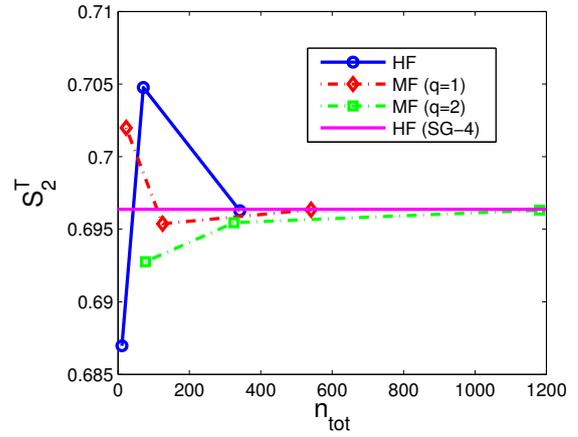}%
\caption{$y_{up}$}%
\label{fig:PARSEC_SP_ST2}%
\end{subfigure}\hfill%
\begin{subfigure}{.45\columnwidth}
\includegraphics[width=1\columnwidth]{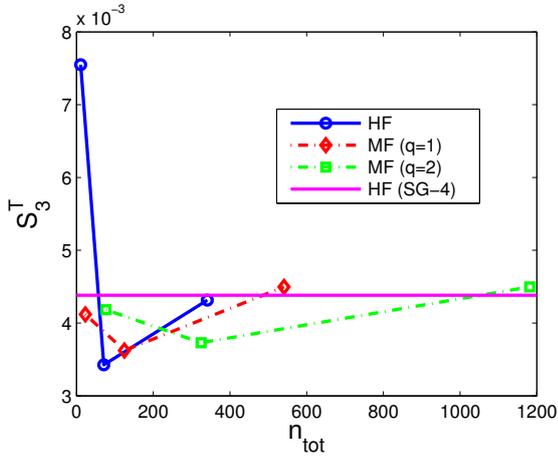}%
\caption{$x_{lo}$}%
\label{fig:PARSEC_SP_ST3}%
\end{subfigure}\hfill%
\begin{subfigure}{.45\columnwidth}
\includegraphics[width=1\columnwidth]{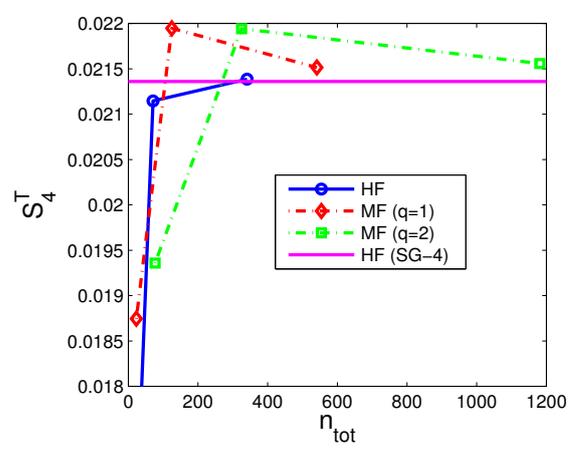}%
\caption{$y_{lo}$}%
\label{fig:PARSEC_SP_ST4}%
\end{subfigure}\hfill%
\begin{subfigure}{.45\columnwidth}
\includegraphics[width=1\columnwidth]{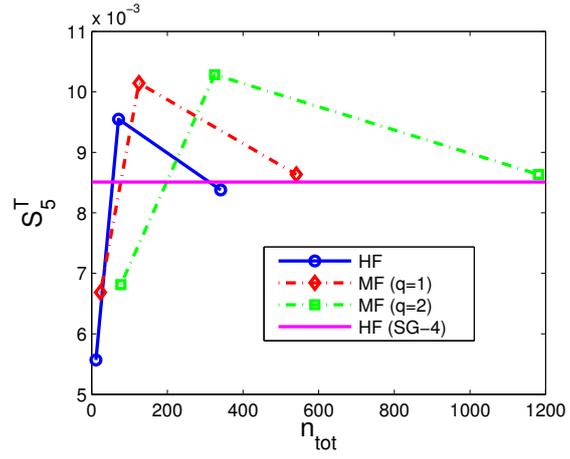}%
\caption{$\alpha_{TE}$}%
\label{fig:PARSEC_SP_ST5}%
\end{subfigure}\hfill%
\caption{Convergence of total Sobol indices for the inviscid transonic airfoil problem. The faster convergence of MF-PCE compared to HF-PCE is especially clear when it is used to approximate the total Sobol indices of $x_{up}$ and $y_{up}$.}
\label{fig:PARSECtot}
\end{figure}

\begin{figure}[H]
\centering
\includegraphics[width=0.5\columnwidth]{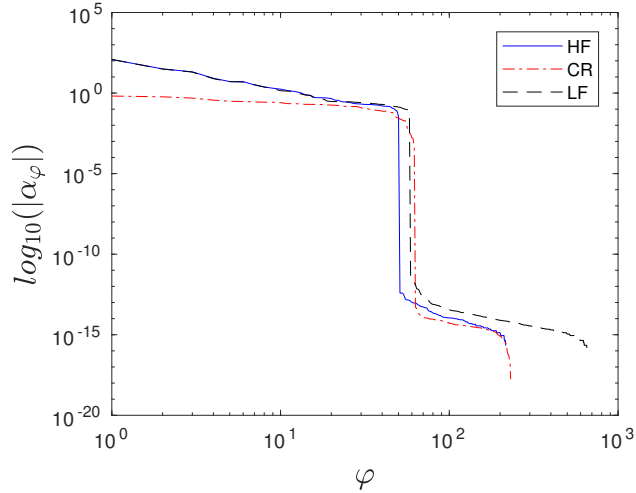}%
\caption{The decay of the correction, HF-, and LF-PCE coefficients for the inviscid transonic airfoil problem. The coefficients of correction PCE evidently decay faster than those of HF and LF-PCE.}
\label{fig:PARSEC_DECAY_SPAM}
\end{figure}

First, convergence analysis of statistical moments and error of total Sobol indices is performed and the results are depicted in Fig.~\ref{fig:PARSECmoments} (the error of all Sobol indices is not shown since the interaction term is small). Shown in the $x-$axis of Fig.~\ref{fig:PARSECmoments} is the actual cost of MF-PCE that includes the cost of the LF function. From the mean and standard deviation perspective, the advantage of utilizing MF-PCE over HF-PCE is not clearly seen. Also, note that there is a jump in computational cost for MF-PCE with $q=2$ when the HF sparse grid level is set to 2. Since increasing the value of $q$ means higher computational cost, one has to be careful if setting the $q$ value too high, especially if the computational cost of the LF function is not negligible. Nonetheless, a clear advantage of MF-PCE is well observed when it is used to estimate the Sobol Indices. Here, MF-PCE with $q=1$ and $q=2$ reach the e$_{\text{T}}$ level below $10^{-3}$ after the equivalent of 125 and 325 HF function evaluations, respectively, where HF-PCE needs 341 function evaluations for the same accuracy; this means that the computational cost is reduced to about 36.66\% and 95.3\% for the MF-PCE method with $q=1$ and $q=2$, respectively. Although the gain from utilizing MF-PCE with $q=2$ seems to be minimal in achieving this accuracy, it reaches the error level of total Sobol indices below $2\times 10^{-3}$ with the equivalent of 77 HF function evaluations. 

Figure~\ref{fig:PARSECtot} shows the convergence of each term of total Sobol indices with respect to the actual cost. The convergence trend is not especially clear for the $x_{lo}$, $y_{lo}$, and $\alpha_{TE}$, since their contribution are relatively small compared to $x_{up}$ and $y_{up}$. We observe that MF-PCE with both $q=1$ and $q=2$ reaches a good accuracy for $x_{up}$ and $y_{up}$ with HF sparse grid level of 2. However, MF-PCE with $q=1$ is more efficient here since it reaches the engineering accuracy with fewer function evaluations than those of MF-PCE with $q=2$ and HF-PCE.

The decay of the correction and LF-PCE coefficients is depicted in Fig.~\ref{fig:PARSEC_DECAY_SPAM}. The plot is prepared by using MF-PCE with correction and LF-PCE sparse grid level of 3 and 4, respectively. By investigating the figure, it is obvious that the correction PCE model is less complex than the LF- and HF-PCE. Thus, the given LF model is appropriate to be used within the MF-PCE framework for this problem. The reason for the similarity between the decay of LF- and HF-PCE coefficients is because that the difference in magnitude between the two functions is small. However, we can see that for polynomial terms with small coefficients (i.e., beyond the 40th term) the two PCEs start to differ. This is where the correction PCE plays its role.

\section{Conclusions}
\label{sec:4}
\noindent An approach for SA using PCE with an MF model is presented in this paper. The goal is to accelerate the SA process when simulations with multiple levels of fidelity are available. The spectral projection-based MF-PCE is used as the method of interest in this paper. The method uses two PCEs: LF and correction, which are combined into a single MF-PCE. SA is directly performed in the post-processing phase by computation of the Sobol indices from the MF-PCE coefficients. 

The proposed method is demonstrated and investigated on some algebraic and engineering test problems. On cases where the correlation of the LF to the HF function is very high (near 1), the Sobol indices obtained from the LF- and HF-PCE are almost the same; revealing that for such cases, the use of LF function is enough to obtain accurate Sobol indices. However, the Sobol indices obtained from LF- and HF-PCE differ greatly if the LF and HF functions are not perfectly correlated. It is also worth noting that one usually wants to compute the statistical moments together with Sobol indices, which means that correction of the LF function is always desired in order to obtain accurate values. The current study shows that MF-PCE can be employed to obtain accurate Sobol indices with lower computational cost than that of HF-PCE. Our investigation shows that MF-PCE is an efficient method for estimating the Sobol indices with a requirement that the LF and HF function are highly correlated with each other. The decrease of error estimation of Sobol indices positively correlated with the decrease of MARE and, especially, r$^{2}$. Investigation on the artificial problems with a hypothetical LF simulations cost suggests that the computational cost ratio between the LF and HF function should be at least lower than $1/4$, in order to ensure the advantage of utilizing MF-PCE over HF-PCE in terms of computational efficiency. In the inviscid transonic airfoil case with five uncertain parameters, the MF approach successfully estimates the Sobol indices with the computational cost of about 36.66\% needed by the HF-PCE method in order to obtain a nearly similar engineering accuracy. 

As for the future work, implementing adaptive sparse grid for MF-PCE is a promising approach to efficiently obtain accurate Sobol indices in high-dimensional problems by focusing on the most important dimensions. Moreover, the Sobol indices themselves can be used to guide the adaptivity. Because such approach is not yet considered, more work is needed to investigate the adaptive MF-PCE approach for SA purpose. Moreover, the use of cost-efficient quadrature such as Clenshaw-Curtis quadrature should also be considered. Investigation on the use of regression-based MF-PCE for SA, which allows the flexibility in sampling location and size, is also a promising research direction.

\section*{Acknowledgement}
Part of this work is funded through Riset KK ITB. Koji Shimoyama was supported in part by the Grant-in-Aid for Scientific Research (B) No.~H1503600 administered by the Japan Society for the Promotion of Science (JSPS).

\nocite{*}
\bibliographystyle{elsarticle-num}
\bibliography{Bibliography}

\begin{thebibliography}{10}
\expandafter\ifx\csname url\endcsname\relax
  \def\url#1{\texttt{#1}}\fi
\expandafter\ifx\csname urlprefix\endcsname\relax\def\urlprefix{URL }\fi
\expandafter\ifx\csname href\endcsname\relax
  \def\href#1#2{#2} \def\path#1{#1}\fi

\bibitem{saltelli2000sensitivity}
A.~Saltelli, K.~Chan, E.~M. Scott, et~al., Sensitivity analysis, Vol.~1, Wiley
  New York, 2000.

\bibitem{efron1981jackknife}
B.~Efron, C.~Stein, The jackknife estimate of variance, The Annals of
  Statistics (1981) 586--596.

\bibitem{sobol1990sensitivity}
I.~M. Sobol', On sensitivity estimation for nonlinear mathematical models,
  Matematicheskoe Modelirovanie 2~(1) (1990) 112--118.

\bibitem{park1991universal}
J.~Park, I.~W. Sandberg, Universal approximation using radial-basis-function
  networks, Neural computation 3~(2) (1991) 246--257.

\bibitem{krige1951statistical}
D.~G. Krige, A statistical approach to some mine valuation and allied problems
  on the {W}itwatersrand, Ph.D. thesis, University of the Witwatersrand (1951).

\bibitem{matheron1963principles}
G.~Matheron, Principles of geostatistics, Economic geology 58~(8) (1963)
  1246--1266.

\bibitem{cressie2015statistics}
N.~Cressie, Statistics for spatial data, John Wiley \& Sons, 2015.

\bibitem{sacks1989design}
J.~Sacks, W.~J. Welch, T.~J. Mitchell, H.~P. Wynn, Design and analysis of
  computer experiments, Statistical science (1989) 409--423.

\bibitem{gandin1965objective}
L.~S. Gandin, R.~Hardin, Objective analysis of meteorological fields, Vol. 242,
  Israel program for scientific translations Jerusalem, 1965.

\bibitem{loeven2007probabilistic}
G.~Loeven, J.~Witteveen, H.~Bijl, Probabilistic collocation: an efficient
  non-intrusive approach for arbitrarily distributed parametric uncertainties,
  in: 45th AIAA Aerospace Sciences Meeting and Exhibit, 2007, p. 317.

\bibitem{knio2001stochastic}
O.~M. Knio, H.~N. Najm, R.~G. Ghanem, et~al., A stochastic projection method
  for fluid flow: I. basic formulation, Journal of computational Physics
  173~(2) (2001) 481--511.

\bibitem{ghiocel2002stochastic}
D.~M. Ghiocel, R.~G. Ghanem, Stochastic finite-element analysis of seismic
  soil-structure interaction, Journal of Engineering Mechanics 128~(1) (2002)
  66--77.

\bibitem{xiu2002wiener}
D.~Xiu, G.~E. Karniadakis, The {W}iener--{A}skey polynomial chaos for
  stochastic differential equations, SIAM journal on scientific computing
  24~(2) (2002) 619--644.

\bibitem{ghanem2003stochastic}
R.~G. Ghanem, P.~D. Spanos, Stochastic finite elements: a spectral approach,
  Courier Corporation, 2003.

\bibitem{das2000cumulative}
P.~Das, Y.~Zheng, Cumulative formation of response surface and its use in
  reliability analysis, Probabilistic Engineering Mechanics 15~(4) (2000)
  309--315.

\bibitem{papadrakakis2002reliability}
M.~Papadrakakis, N.~D. Lagaros, Reliability-based structural optimization using
  neural networks and monte carlo simulation, Computer methods in applied
  mechanics and engineering 191~(32) (2002) 3491--3507.

\bibitem{kaymaz2005application}
I.~Kaymaz, Application of kriging method to structural reliability problems,
  Structural Safety 27~(2) (2005) 133--151.

\bibitem{dubourg2011metamodel}
V.~Dubourg, F.~Deheeger, B.~Sudret, Metamodel-based importance sampling for the
  simulation of rare events, Applications of Statistics and Probability in
  Civil Engineering 26 (2011) 192.

\bibitem{bichon2011efficient}
B.~J. Bichon, J.~M. McFarland, S.~Mahadevan, Efficient surrogate models for
  reliability analysis of systems with multiple failure modes, Reliability
  Engineering \& System Safety 96~(10) (2011) 1386--1395.

\bibitem{balesdent2013kriging}
M.~Balesdent, J.~Morio, J.~Marzat, Kriging-based adaptive importance sampling
  algorithms for rare event estimation, Structural Safety 44 (2013) 1--10.

\bibitem{bichon2008efficient}
B.~J. Bichon, M.~S. Eldred, L.~P. Swiler, S.~Mahadevan, J.~M. McFarland,
  Efficient global reliability analysis for nonlinear implicit performance
  functions, AIAA J 46~(10) (2008) 2459--2468.

\bibitem{echard2011ak}
B.~Echard, N.~Gayton, M.~Lemaire, Ak-mcs: an active learning reliability method
  combining kriging and monte carlo simulation, Structural Safety 33~(2) (2011)
  145--154.

\bibitem{huang2016assessing}
X.~Huang, J.~Chen, H.~Zhu, Assessing small failure probabilities by ak--ss: an
  active learning method combining kriging and subset simulation, Structural
  Safety 59 (2016) 86--95.

\bibitem{choi2004structural}
S.-K. Choi, R.~V. Grandhi, R.~A. Canfield, Structural reliability under
  non-gaussian stochastic behavior, Computers \& structures 82~(13) (2004)
  1113--1121.

\bibitem{berveiller2006stochastic}
M.~Berveiller, B.~Sudret, M.~Lemaire, Stochastic finite element: a non
  intrusive approach by regression, European Journal of Computational
  Mechanics/Revue Europ{\'e}enne de M{\'e}canique Num{\'e}rique 15~(1-3) (2006)
  81--92.

\bibitem{hu2011adaptive}
C.~Hu, B.~D. Youn, Adaptive-sparse polynomial chaos expansion for reliability
  analysis and design of complex engineering systems, Structural and
  Multidisciplinary Optimization 43~(3) (2011) 419--442.

\bibitem{sudret2013response}
B.~Sudret, G.~Blatman, M.~Berveiller, Response surfaces based on polynomial
  chaos expansions, Construction reliability: safety, variability and
  sustainability (2013) 147--167.

\bibitem{schobi2016rare}
R.~Sch{\"o}bi, B.~Sudret, S.~Marelli, Rare event estimation using
  polynomial-chaos kriging, ASCE-ASME Journal of Risk and Uncertainty in
  Engineering Systems, Part A: Civil Engineering 3~(2) (2016) D4016002.

\bibitem{jin2002sequential}
R.~Jin, W.~Chen, A.~Sudjianto, On sequential sampling for global metamodeling
  in engineering design, in: Proceedings of DETC, Vol.~2, 2002, pp. 539--548.

\bibitem{bilionis2012multi}
I.~Bilionis, N.~Zabaras, Multi-output local gaussian process regression:
  Applications to uncertainty quantification, Journal of Computational Physics
  231~(17) (2012) 5718--5746.

\bibitem{shimoyama2013dynamic}
K.~Shimoyama, S.~Kawai, J.~J. Alonso, Dynamic adaptive sampling based on
  kriging surrogate models for efficient uncertainty quantification, in: 15th
  AIAA Non-Deterministic Approaches Conference, 2013, pp. 2013--1470.

\bibitem{dwight2009efficient}
R.~P. Dwight, Z.-H. Han, Efficient uncertainty quantification using
  gradient-enhanced kriging, AIAA paper 2276 (2009) 2009.

\bibitem{rumpfkeil2011dynamic}
M.~P. Rumpfkeil, W.~Yamazaki, D.~J. Mavriplis, A dynamic sampling method for
  kriging and cokriging surrogate models, in: 49th AIAA Aerospace Sciences
  Meeting including the New Horizons Forum and Aerospace Exposition. American
  Institute of Aeronautics et Astronautics (cf. p. 162), 2011.

\bibitem{wang2010high}
Q.~Wang, P.~Moin, G.~Iaccarino, A high order multivariate approximation scheme
  for scattered data sets, Journal of Computational Physics 229~(18) (2010)
  6343--6361.

\bibitem{boopathy2014unified}
K.~Boopathy, M.~P. Rumpfkeil, Unified framework for training point selection
  and error estimation for surrogate models, Aiaa Journal.

\bibitem{sudret2008global}
B.~Sudret, Global sensitivity analysis using polynomial chaos expansions,
  Reliability Engineering \& System Safety 93~(7) (2008) 964--979.

\bibitem{crestaux2009polynomial}
T.~Crestaux, O.~Le~Ma{\i}ˆtre, J.-M. Martinez, Polynomial chaos expansion for
  sensitivity analysis, Reliability Engineering \& System Safety 94~(7) (2009)
  1161--1172.

\bibitem{morris1993bayesian}
M.~D. Morris, T.~J. Mitchell, D.~Ylvisaker, Bayesian design and analysis of
  computer experiments: use of derivatives in surface prediction, Technometrics
  35~(3) (1993) 243--255.

\bibitem{liu2003development}
W.~Liu, Development of gradient-enhanced kriging approximations for
  multidisciplinary design optimization, 2003.

\bibitem{de2014improvements}
J.~H. de~Baar, R.~P. Dwight, H.~Bijl, Improvements to gradient-enhanced kriging
  using a bayesian interpretation, International Journal for Uncertainty
  Quantification 4~(3).

\bibitem{de2015exploiting}
J.~de~Baar, T.~Scholcz, R.~Dwight, Exploiting adjoint derivatives in
  high-dimensional metamodels, AIAA Journal 53~(5) (2015) 1391--1395.

\bibitem{peng2016polynomial}
J.~Peng, J.~Hampton, A.~Doostan, On polynomial chaos expansion via
  gradient-enhanced ℓ 1-minimization, Journal of Computational Physics.

\bibitem{de2015gradient}
J.~H. de~Baar, B.~Harding, A gradient-enhanced sparse grid algorithm for
  uncertainty quantification, International Journal for Uncertainty
  Quantification 5~(5).

\bibitem{wiener1938homogeneous}
N.~Wiener, The homogeneous chaos, American Journal of Mathematics 60~(4) (1938)
  897--936.

\bibitem{choi2004polynomial}
S.-K. Choi, R.~V. Grandhi, R.~A. Canfield, C.~L. Pettit, Polynomial chaos
  expansion with latin hypercube sampling for estimating response variability,
  AIAA journal 42~(6) (2004) 1191--1198.

\bibitem{eldred2009comparison}
M.~S. Eldred, J.~Burkardt, Comparison of non-intrusive polynomial chaos and
  stochastic collocation methods for uncertainty quantification, in:
  Proceedings of the 47th AIAA Aerospace Sciences Meeting, no. 2009-0976, AIAA,
  Orlando, FL, 2009.

\bibitem{blatman2010efficient}
G.~Blatman, B.~Sudret, Efficient computation of global sensitivity indices
  using sparse polynomial chaos expansions, Reliability Engineering \& System
  Safety 95~(11) (2010) 1216--1229.

\bibitem{blatman2011adaptive}
G.~Blatman, B.~Sudret, Adaptive sparse polynomial chaos expansion based on
  least angle regression, Journal of Computational Physics 230~(6) (2011)
  2345--2367.

\bibitem{doostan2011non}
A.~Doostan, H.~Owhadi, A non-adapted sparse approximation of {PDE}s with
  stochastic inputs, Journal of Computational Physics 230~(8) (2011)
  3015--3034.

\bibitem{jakeman2015enhancing}
J.~D. Jakeman, M.~S. Eldred, K.~Sargsyan, Enhancing ℓ1-minimization estimates
  of polynomial chaos expansions using basis selection, Journal of
  Computational Physics 289 (2015) 18--34.

\bibitem{smolyak1960interpolation}
S.~Smolyak, Interpolation and quadrature formulas for the classes wa s and ea
  s, in: Dokl. Akad. Nauk SSSR, Vol. 131, 1960, pp. 1028--1031.

\bibitem{gerstner1998numerical}
T.~Gerstner, M.~Griebel, Numerical integration using sparse grids, Numerical
  algorithms 18~(3) (1998) 209--232.

\bibitem{bungartz2004sparse}
H.-J. Bungartz, M.~Griebel, Sparse grids, Acta numerica 13 (2004) 147--269.

\bibitem{garcke2012sparse}
J.~Garcke, Sparse grids in a nutshell, in: Sparse grids and applications,
  Springer, 2012, pp. 57--80.

\bibitem{smolyak1963quadrature}
S.~A. Smolyak, Quadrature and interpolation formulas for tensor products of
  certain classes of functions, in: Dokl. Akad. Nauk SSSR, Vol.~4, 1963, p.
  123.

\bibitem{xiu2005high}
D.~Xiu, J.~S. Hesthaven, High-order collocation methods for differential
  equations with random inputs, SIAM Journal on Scientific Computing 27~(3)
  (2005) 1118--1139.

\bibitem{constantine2012sparse}
P.~G. Constantine, M.~S. Eldred, E.~T. Phipps, Sparse pseudospectral
  approximation method, Computer Methods in Applied Mechanics and Engineering
  229 (2012) 1--12.

\bibitem{buzzard2012global}
G.~T. Buzzard, Global sensitivity analysis using sparse grid interpolation and
  polynomial chaos, Reliability Engineering \& System Safety 107 (2012) 82--89.

\bibitem{buzzard2011variance}
G.~T. Buzzard, D.~Xiu, Variance-based global sensitivity analysis via
  sparse-grid interpolation and cubature, Communications in Computational
  Physics 9~(03) (2011) 542--567.

\bibitem{garcia2014global}
O.~Garcia-Cabrejo, A.~Valocchi, Global sensitivity analysis for multivariate
  output using polynomial chaos expansion, Reliability Engineering \& System
  Safety 126 (2014) 25--36.

\bibitem{kennedy2000predicting}
M.~C. Kennedy, A.~O'Hagan, Predicting the output from a complex computer code
  when fast approximations are available, Biometrika 87~(1) (2000) 1--13.

\bibitem{forrester2007multi}
A.~I. Forrester, A.~S{\'o}bester, A.~J. Keane, Multi-fidelity optimization via
  surrogate modelling, in: Proceedings of the {R}oyal {S}ociety of {L}ondon a:
  mathematical, physical and engineering sciences, Vol. 463, The Royal Society,
  2007, pp. 3251--3269.

\bibitem{bandler2004space}
J.~W. Bandler, Q.~S. Cheng, S.~Dakroury, A.~S. Mohamed, M.~H. Bakr, K.~Madsen,
  J.~S{\o}ndergaard, et~al., Space mapping: the state of the art, Microwave
  Theory and Techniques, IEEE Transactions on 52~(1) (2004) 337--361.

\bibitem{shah2015multi}
H.~Shah, S.~Hosder, S.~Koziel, Y.~A. Tesfahunegn, L.~Leifsson, Multi-fidelity
  robust aerodynamic design optimization under mixed uncertainty, Aerospace
  Science and Technology 45 (2015) 17--29.

\bibitem{giles2008multilevel}
M.~B. Giles, Multilevel {M}onte {C}arlo path simulation, Operations Research
  56~(3) (2008) 607--617.

\bibitem{barth2011multi}
A.~Barth, C.~Schwab, N.~Zollinger, Multi-level {M}onte {C}arlo finite element
  method for elliptic {PDE}s with stochastic coefficients, Numerische
  Mathematik 119~(1) (2011) 123--161.

\bibitem{ng2014multifidelity}
L.~W. Ng, K.~E. Willcox, Multifidelity approaches for optimization under
  uncertainty, International Journal for Numerical Methods in Engineering
  100~(10) (2014) 746--772.

\bibitem{de2015uncertainty}
J.~de~Baar, S.~Roberts, R.~Dwight, B.~Mallol, Uncertainty quantification for a
  sailing yacht hull, using multi-fidelity kriging, Computers \& Fluids 123
  (2015) 185--201.

\bibitem{ng2012multifidelity}
L.~W.-T. Ng, M.~Eldred, Multifidelity uncertainty quantification using
  nonintrusive polynomial chaos and stochastic collocation, in: Proceedings of
  the 14th AIAA Non-Deterministic Approaches Conference, number AIAA-2012-1852,
  Honolulu, HI, Vol.~45, 2012.

\bibitem{palar2015decomposition}
P.~S. Palar, T.~Tsuchiya, G.~Parks, Decomposition-based evolutionary
  aerodynamic robust optimization with multi-fidelity point collocation
  non-intrusive polynomial chaos, in: 17th AIAA Non-Deterministic Approaches
  Conference, 2015, p. 1377.

\bibitem{palar2016multi}
P.~S. Palar, T.~Tsuchiya, G.~T. Parks, Multi-fidelity non-intrusive polynomial
  chaos based on regression, Computer Methods in Applied Mechanics and
  Engineering 305 (2016) 579--606.

\bibitem{narayan2014stochastic}
A.~Narayan, C.~Gittelson, D.~Xiu, A stochastic collocation algorithm with
  multifidelity models, SIAM Journal on Scientific Computing 36~(2) (2014)
  A495--A521.

\bibitem{le2014bayesian}
L.~Le~Gratiet, C.~Cannamela, B.~Iooss, A {B}ayesian approach for global
  sensitivity analysis of (multifidelity) computer codes, SIAM/ASA Journal on
  Uncertainty Quantification 2~(1) (2014) 336--363.

\bibitem{homma1996importance}
T.~Homma, A.~Saltelli, Importance measures in global sensitivity analysis of
  nonlinear models, Reliability Engineering \& System Safety 52~(1) (1996)
  1--17.

\bibitem{xiu2003modeling}
D.~Xiu, G.~E. Karniadakis, Modeling uncertainty in flow simulations via
  generalized polynomial chaos, Journal of computational physics 187~(1) (2003)
  137--167.

\bibitem{askey1985some}
R.~Askey, J.~A. Wilson, Some basic hypergeometric orthogonal polynomials that
  generalize Jacobi polynomials, Vol. 319, American Mathematical Soc., 1985.

\bibitem{griebel1990combination}
M.~Griebel, M.~Schneider, C.~Zenger, A combination technique for the solution
  of sparse grid problems, Technische Universit{\"a}t, 1990.

\bibitem{toal2015some}
D.~J. Toal, Some considerations regarding the use of multi-fidelity kriging in
  the construction of surrogate models, Structural and Multidisciplinary
  Optimization 51~(6) (2015) 1223--1245.

\bibitem{xiong2013sequential}
S.~Xiong, P.~Z. Qian, C.~J. Wu, Sequential design and analysis of high-accuracy
  and low-accuracy computer codes, Technometrics 55~(1) (2013) 37--46.

\bibitem{ishigami1990importance}
T.~Ishigami, T.~Homma, An importance quantification technique in uncertainty
  analysis for computer models, in: Uncertainty Modeling and Analysis, 1990.
  Proceedings., First International Symposium on, IEEE, 1990, pp. 398--403.

\bibitem{palacios2013stanford}
F.~Palacios, M.~R. Colonno, A.~C. Aranake, A.~Campos, S.~R. Copeland, T.~D.
  Economon, A.~K. Lonkar, T.~W. Lukaczyk, T.~W. Taylor, J.~J. Alonso, Stanford
  university unstructured ({SU}$^{2}$): An open-source integrated computational
  environment for multi-physics simulation and design, AIAA Paper 287 (2013)
  2013.

\bibitem{sobieczky1999parametric}
H.~Sobieczky, Parametric airfoils and wings, in: Recent Development of
  Aerodynamic Design Methodologies, Springer, 1999, pp. 71--87.

\bibitem{forrester2006optimization}
A.~I. Forrester, N.~W. Bressloff, A.~J. Keane, Optimization using surrogate
  models and partially converged computational fluid dynamics simulations, in:
  Proceedings of the Royal Society of London A: Mathematical, Physical and
  Engineering Sciences, Vol. 462, The Royal Society, 2006, pp. 2177--2204.

\bibitem{branke2017efficient}
J.~Branke, M.~Asafuddoula, K.~S. Bhattacharjee, T.~Ray, Efficient use of
  partially converged simulations in evolutionary optimization, IEEE
  Transactions on Evolutionary Computation 21~(1) (2017) 52--64.

\bibitem{courrier2014use}
N.~Courrier, P.-A. Boucard, B.~Soulier, The use of partially converged
  simulations in building surrogate models, Advances in Engineering Software 67
  (2014) 186--197.

\bibitem{liu2017survey}
H.~Liu, Y.-S. Ong, J.~Cai, A survey of adaptive sampling for global
  metamodeling in support of simulation-based complex engineering design,
  Structural and Multidisciplinary Optimization (2017) 1--24.

\bibitem{wang2013application}
P.~Wang, Z.~Lu, Z.~Tang, An application of the {K}riging method in global
  sensitivity analysis with parameter uncertainty, Applied Mathematical
  Modelling 37~(9) (2013) 6543--6555.

\end{thebibliography}

\end{document}